\documentclass[a4paper,preprint,showpacs,superscriptaddress]{revtex4}
\usepackage[version=3]{mhchem}
\usepackage{float,fancyhdr,amsmath,graphicx,subfig,epsfig,color,setspace,url}
\usepackage{appendix}
\begin{document}

\title{Dissociative Electron Attachment to Polyatomic Molecules - II : Hydrogen Sulphide}
\author{N. Bhargava Ram}
\email[]{nbhargavaram@tifr.res.in}
\affiliation{Tata Institute of Fundamental Research, Mumbai 400005, India}

\author{E. Krishnakumar}
\email[]{ekkumar@tifr.res.in}
\affiliation{Tata Institute of Fundamental Research,  Mumbai 400005, India}

\begin{abstract}
In the present paper, we report the details of the kinetic energy and angular distributions of the \ce{H-}, \ce{S-}/\ce{SH-} fragment ions produced due to dissociative electron attachment to \ce{H2S} molecules at resonances peaking at 2.2 eV, 5.2 eV, 7.5 eV and 10 eV.
\end{abstract}

\maketitle

\section{Introduction}

As described in the preceding paper on DEA to water, electron attachment to water is beset with complex dissociation dynamics deviating from axial recoil approximation, especially at the second and third resonances (\ce{^{2}A1} and \ce{^{2}B2} resonances). It was a logical extension to do similar measurements on \ce{H2S} - a molecule similar to \ce{H2O} and compare the processes due to electron attachment in these two molecules. Sulphur and Oxygen belong to the same group in the periodic table and hence \ce{H2S} has an iso-electronic valence shell similar to \ce{H2O} and similar hierarchy of molecular orbitals. The ground state electronic configuration of neutral \ce{H2S} is \ce{1a1^{2} 2a1^{2} 1b2^{2} 3a1^{2} 1b1^{2} 4a1^{2} 2b2^{2} 5a1^{2} 2b1^{2} -> ^{1}A1} symmetry (\ce{C_{2v}} geometry). The vacant orbitals thereafter are \ce{6a1} and \ce{3b2} respectively. Total ion measurements in \ce{H2S} following electron attachment have shown four resonant peaks at 2.4 eV, 5.2 eV, 7.5 eV and at about 10 eV respectively \cite{c4azria}. The peak at 2.4 eV is due to a shape resonance where the incident electron is attached in the ground state configuration of the neutral \ce{H2S} molecule. The three higher energy peaks are due to \ce{H2S^{-*}} states with the extra electron attached to the electronically excited states of the neutral \ce{H2S}. The shape resonance seen at low energy in \ce{H2S} is not seen in \ce{H2O}. The \ce{H2S^{-*}} anion state may decay by dissociation through one of the dissociation channels listed in Table \ref{tab4.1} along with their appearance energies.

\begin{table}[!h]
\caption{Various dissociation channels on electron attachment to \ce{H2S} with threshold}
\begin{center}
\begin{tabular}{ccccll}
\hline
\\
&&&& Dissociation channel & Threshold \\
\\
\hline
\\
\ce{H2S} + \ce{e-} & $\rightarrow$ & \ce{H2S^{-*}} & $\rightarrow$ & \ce{H-} + SH (X $^{2}\Pi$) & 3.15 eV \\
\\
&&&& \ce{H-} + SH (A $^{2}\Sigma)$ & 6.89 eV \\
\\ 
&&&& \ce{H-} + H + S & 6.75 eV \\
\\
&&&& \ce{S-} + \ce{H2} & 1.06 eV \\
\\ 
&&&& \ce{S-} + H + H & 5.4 eV \\
\\
&&&& \ce{SH-}($^{1}\Sigma^{+}$) + H & 1.58 eV \\
\\
\hline
\end{tabular}
\end{center}
\label{tab4.1}
\end{table}

One of the earliest studies on DEA to \ce{H2S} was reported by Fiquet-Fayard et al. \cite{c4fiquet} and Azria et al \cite{c4azria2}. Fiquet-Fayard et al. \cite{c4fiquet} determined the cross section of the \ce{SH}  fragment formation due to DEA in \ce{H2S} for the shape resonance occurring at electron energies close to 2 eV and also studied isotope effects in the deuterated counter parts such as \ce{HDS} and \ce{D2S}. Azria et al. \cite{c4azria2} measured the ion yield spectrum at the resonances peaking at 2.2 eV, 5.35 eV, 8 eV and 10 eV and identified the fragment ions and the isotope effects. The peak 10 eV was identified as mostly due to \ce{S-} arising from a three-body fragmentation. In 1979, Azria et al. \cite{c4azria} reported kinetic energy and angular distributions of \ce{H-} ions at 5.2 eV and 7.5 eV. They identified these resonances as \ce{^{2}B1} and \ce{^{2}A1} respectively based on calculations extending the O'Malley and Taylor \cite{c4omalleytaylor} results for a \ce{C_{2v}} point group molecule.  Recently, Abouaf and Teillet-Billy \cite{c4abouaf} made high resolution ion yield curve measurements on \ce{S-} and \ce{SH-} fragments up to 12 eV. They observed the presence of both \ce{S-} and \ce{SH-} at the 2.4 eV resonance. While the \ce{S-} channel also shows a peak at 6 eV and a broad intense peak between 8 and 12 eV, the \ce{SH-} fragment is present only at the 2.4 eV resonance. Electron scattering studies by Rohr \cite{c4rohr} supported the assignment of \ce{^{2}A1} symmetry to the 2.4 eV resonance, though theoretical calculations \cite{c4jain,c4gianturco,c4varella,c4gupta,c4gulley} indicate this as a \ce{^{2}B2} state. For the resonance at 5.5 eV, Haxton et al. \cite{c4h4} applied the local complex potential model to \ce{H2S} just as they did to \ce{H2O}. Considering a \ce{^{2}B1} resonance, they were able to take into account the overall angular behaviour for the \ce{H-} + HS ($^{2}\Pi$, $\nu$ = 0) process at 5.5 eV, and even have a qualitative agreement for the behaviour of the process \ce{H-} +HS ($^{2}\Pi$, $\nu$ = 0) with the data of Azria et al \cite{c4azria}. 

As described above, there have been a few measurements to characterize the resonances at 2.2 eV, 5.2 eV and 7.5 eV. However, so far there has been no kinetic energy and angular distribution measurements of the fragment ions from the resonance process at 10 eV. Moreover, like in the case of other molecules, it is expected that more details of the DEA process could be obtained using the ion momentum imaging technique as compared to the conventional techniques. In this context, our measurements on \ce{H2S} covering all the resonances using the velocity map imaging technique assume significance in unravelling the dissociation dynamics therein and understanding the similarities / dissimilarities vis-a-vis \ce{H2O}.

\section{Results and Discussion}

\begin{figure}[!htp]
\centering
\includegraphics[width=0.5\columnwidth]{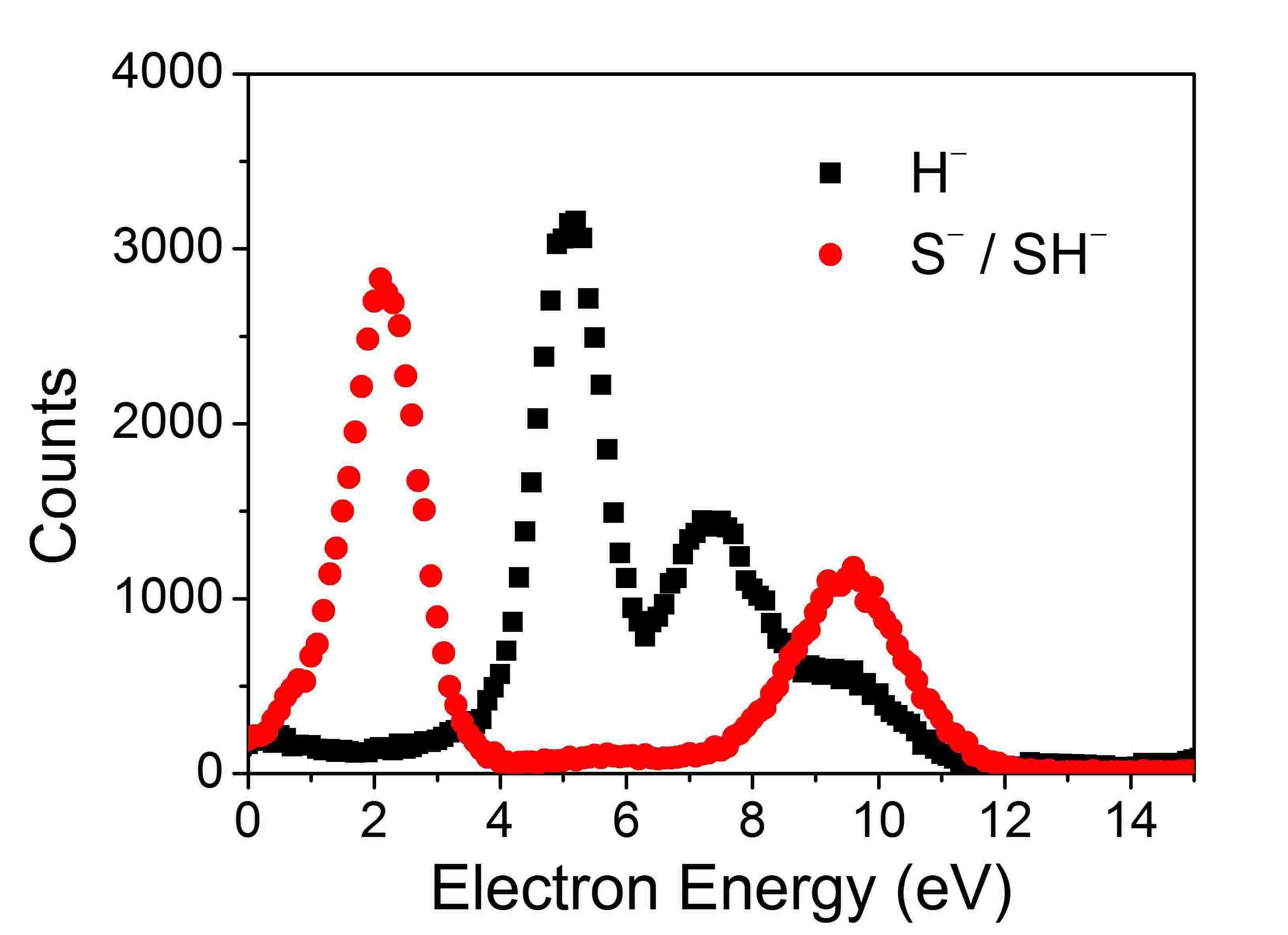}
\caption{Ion yield curves of \ce{H-} and \ce{S-}/\ce{SH-} ions from DEA to \ce{H2S}. \ce{S-} and \ce{SH-} are unresolved in the measurement. Figure not to scale.}
\label{fig4.1}
\end{figure} 

\begin{figure}[!h]
\centering
   \subfloat[$H^{-}$ at 4.2 eV]{\includegraphics[width=0.3\columnwidth]{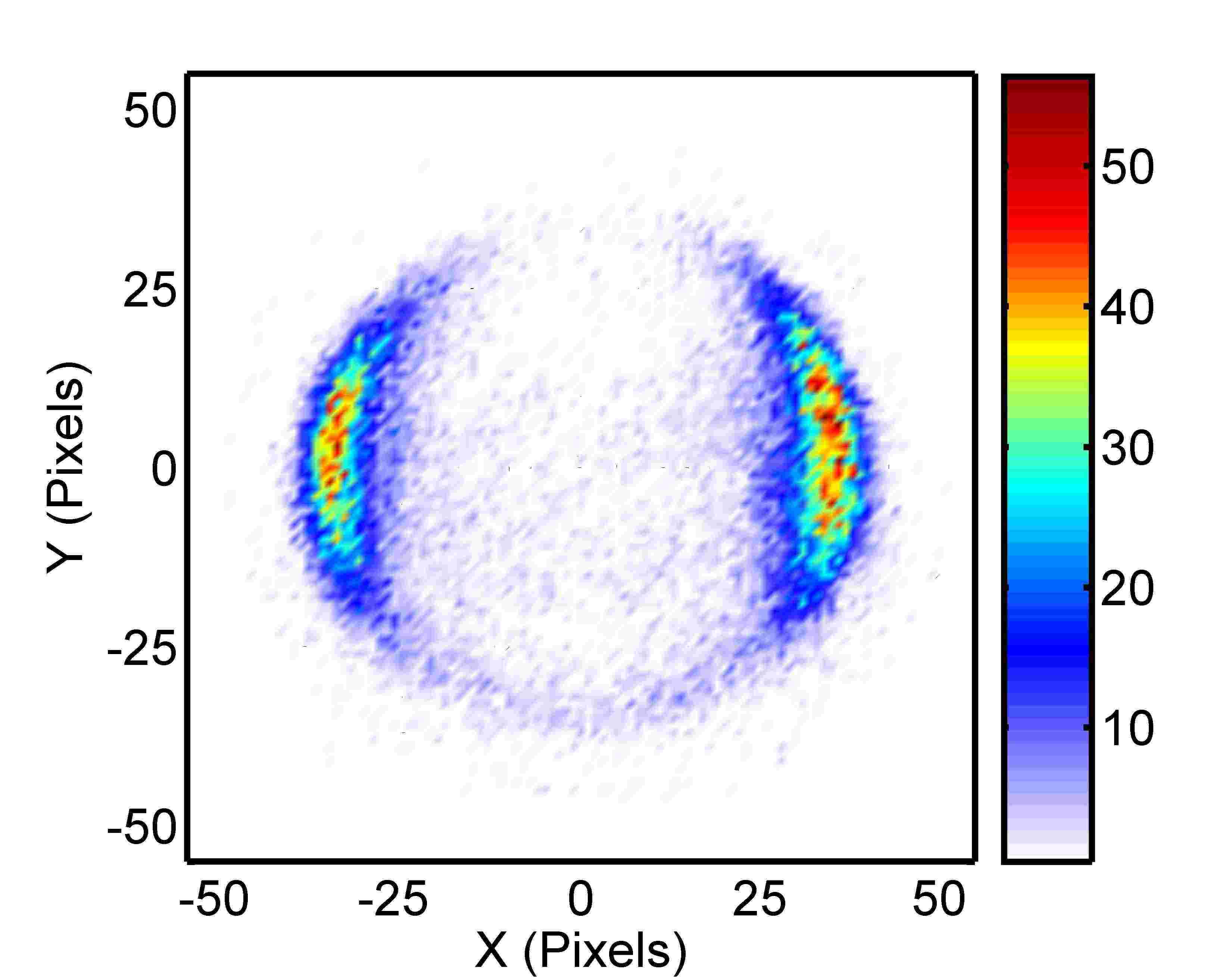}}
  \subfloat[$H^{-}$ at 5.2 eV]{\includegraphics[width=0.3\columnwidth]{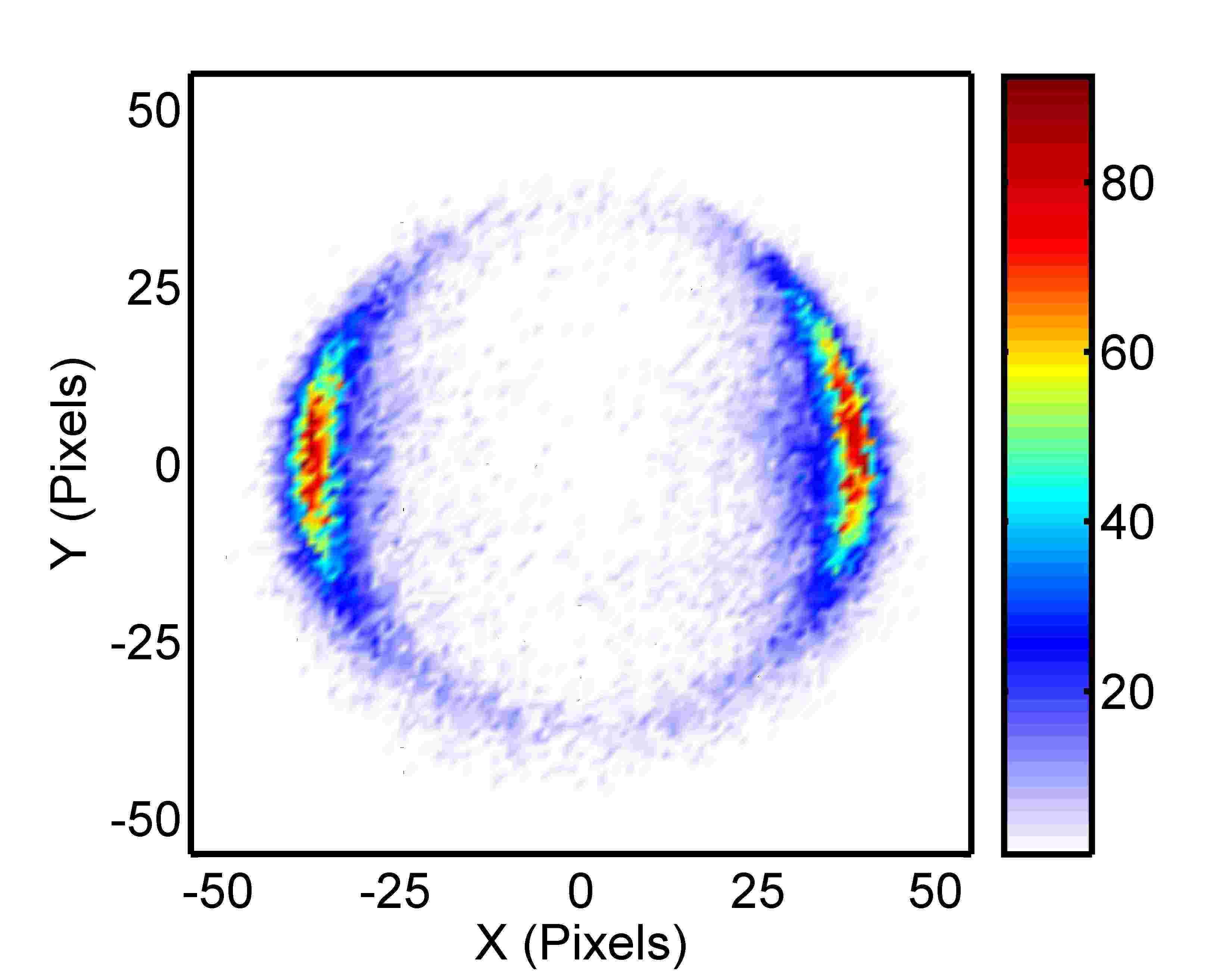}}
  \subfloat[$H^{-}$ at 6.0 eV]{\includegraphics[width=0.3\columnwidth]{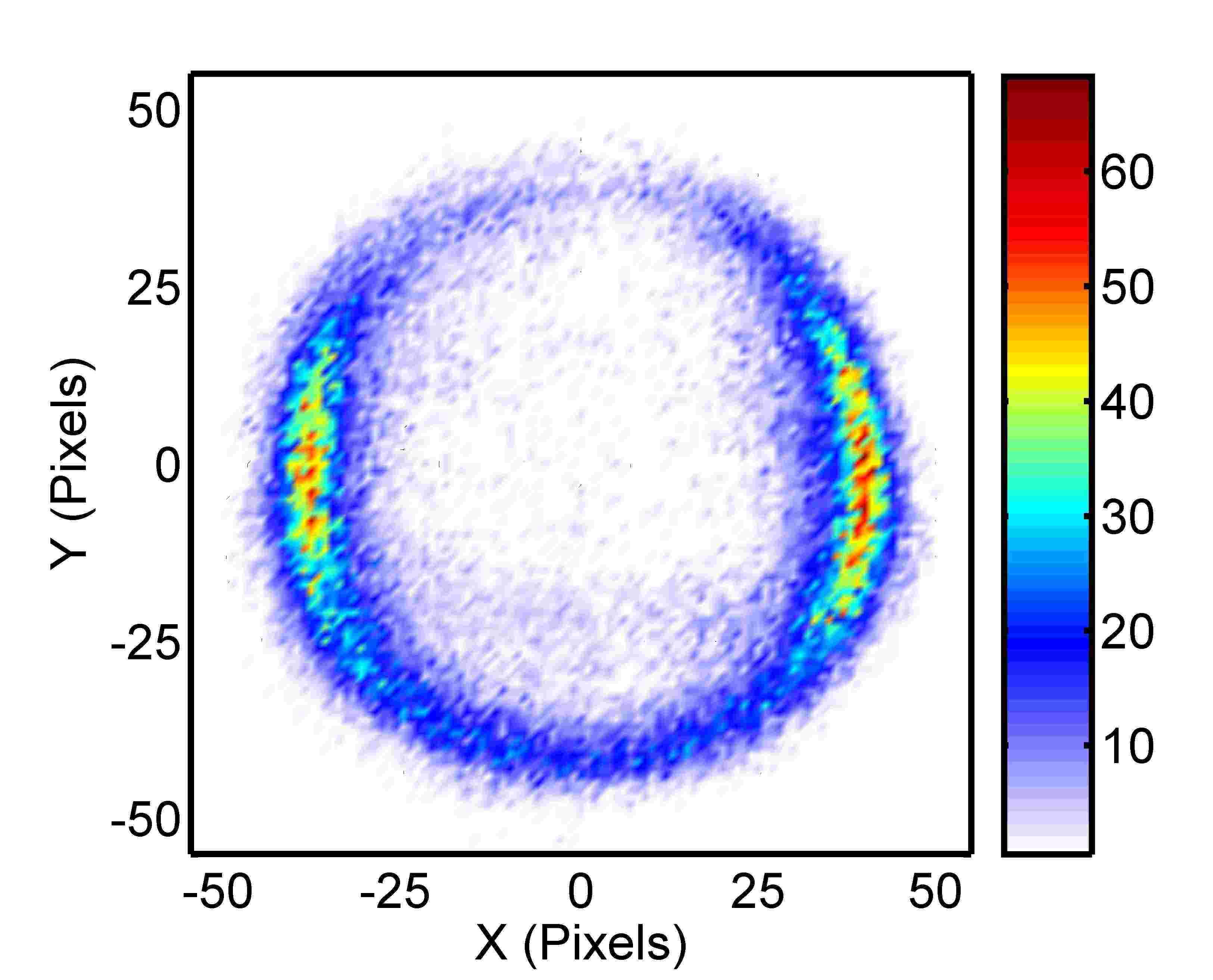}}\\
  \subfloat[$H^{-}$ at 6.8 eV]{\includegraphics[width=0.3\columnwidth]{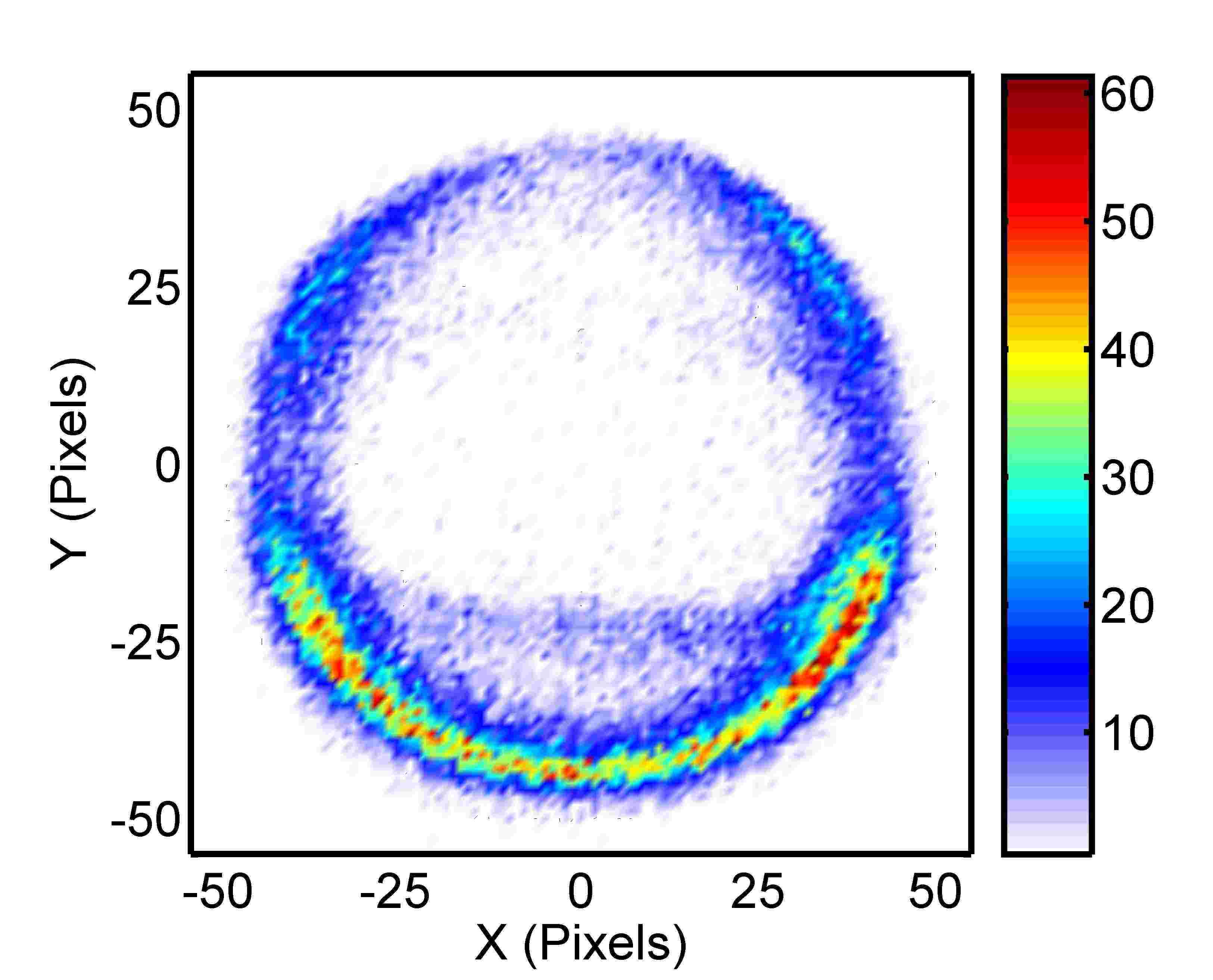}}
  \subfloat[$H^{-}$ at 7.5 eV]{\includegraphics[width=0.3\columnwidth]{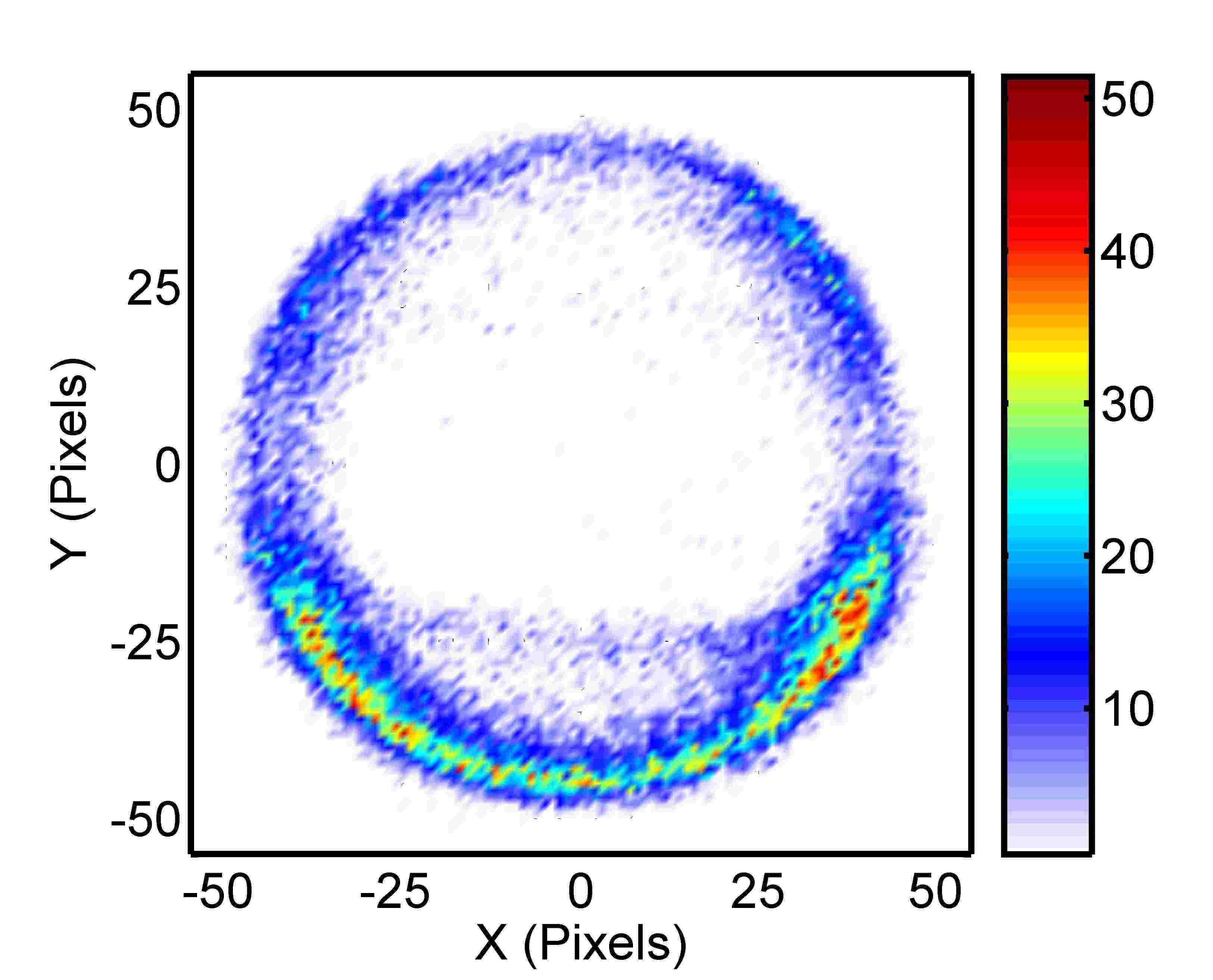}}
   \subfloat[$H^{-}$ at 8.2 eV]{\includegraphics[width=0.3\columnwidth]{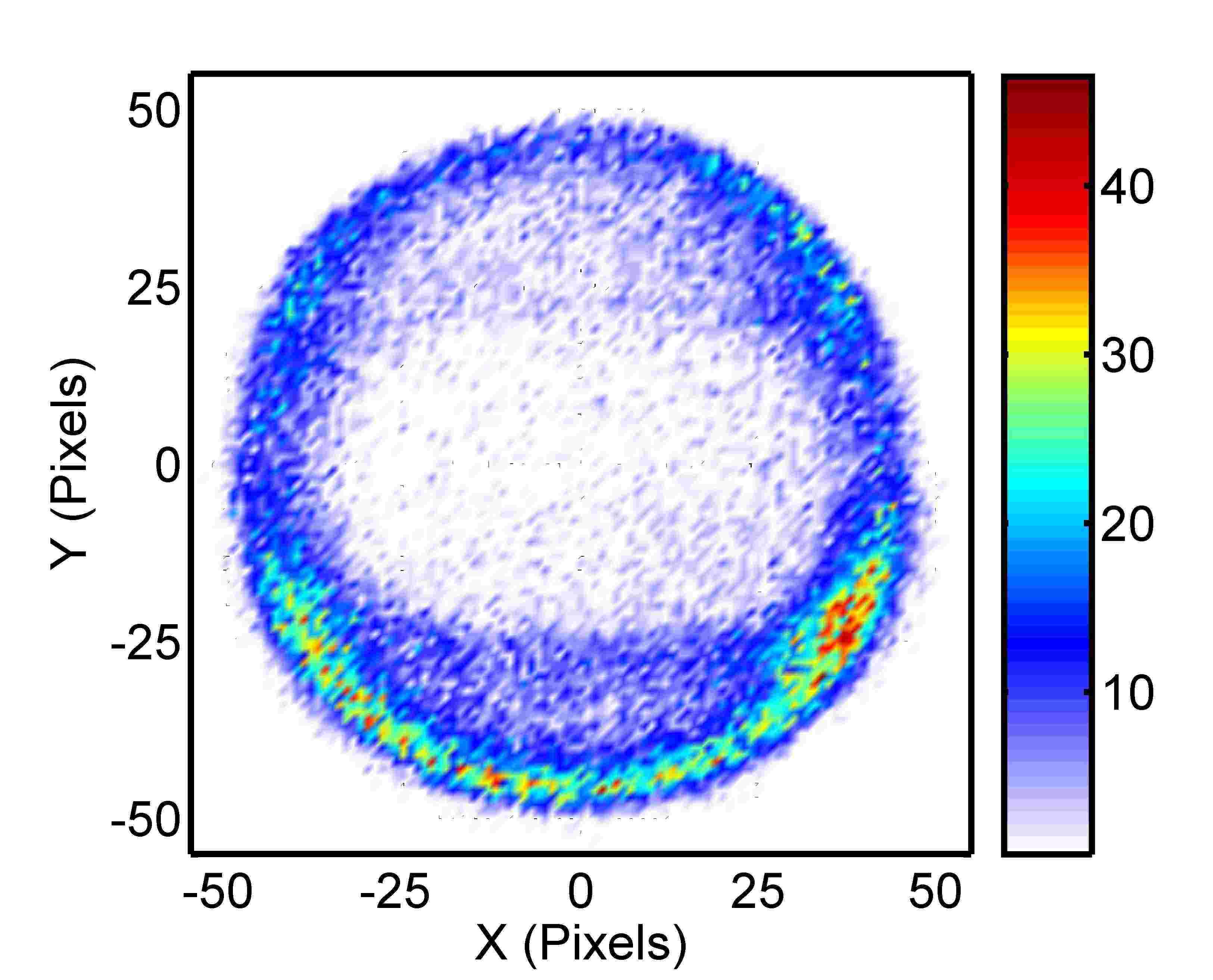}}\\
  \subfloat[$H^{-}$ at 9.0 eV]{\includegraphics[width=0.3\columnwidth]{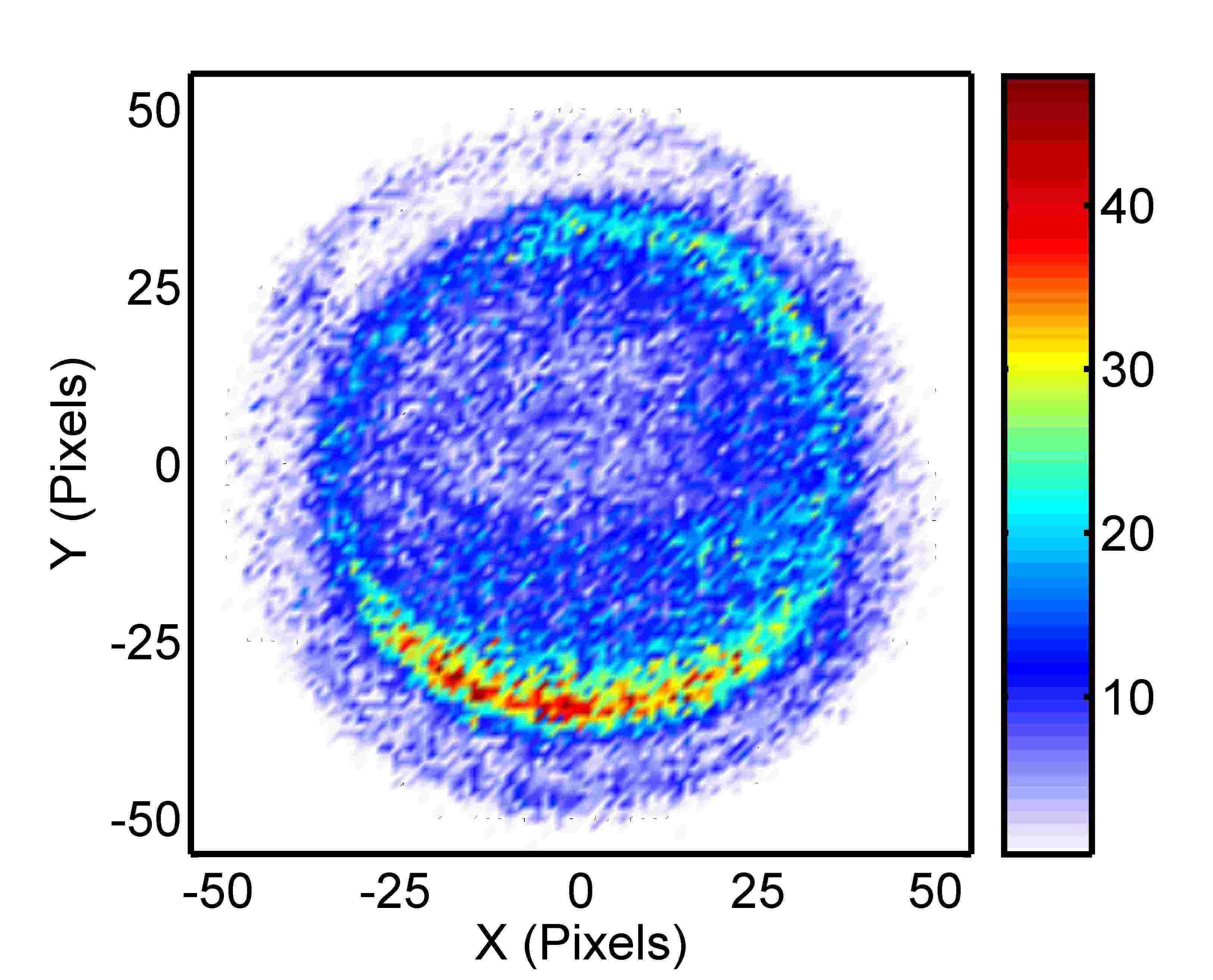}}
  \subfloat[$H^{-}$ at 9.6 eV]{\includegraphics[width=0.3\columnwidth]{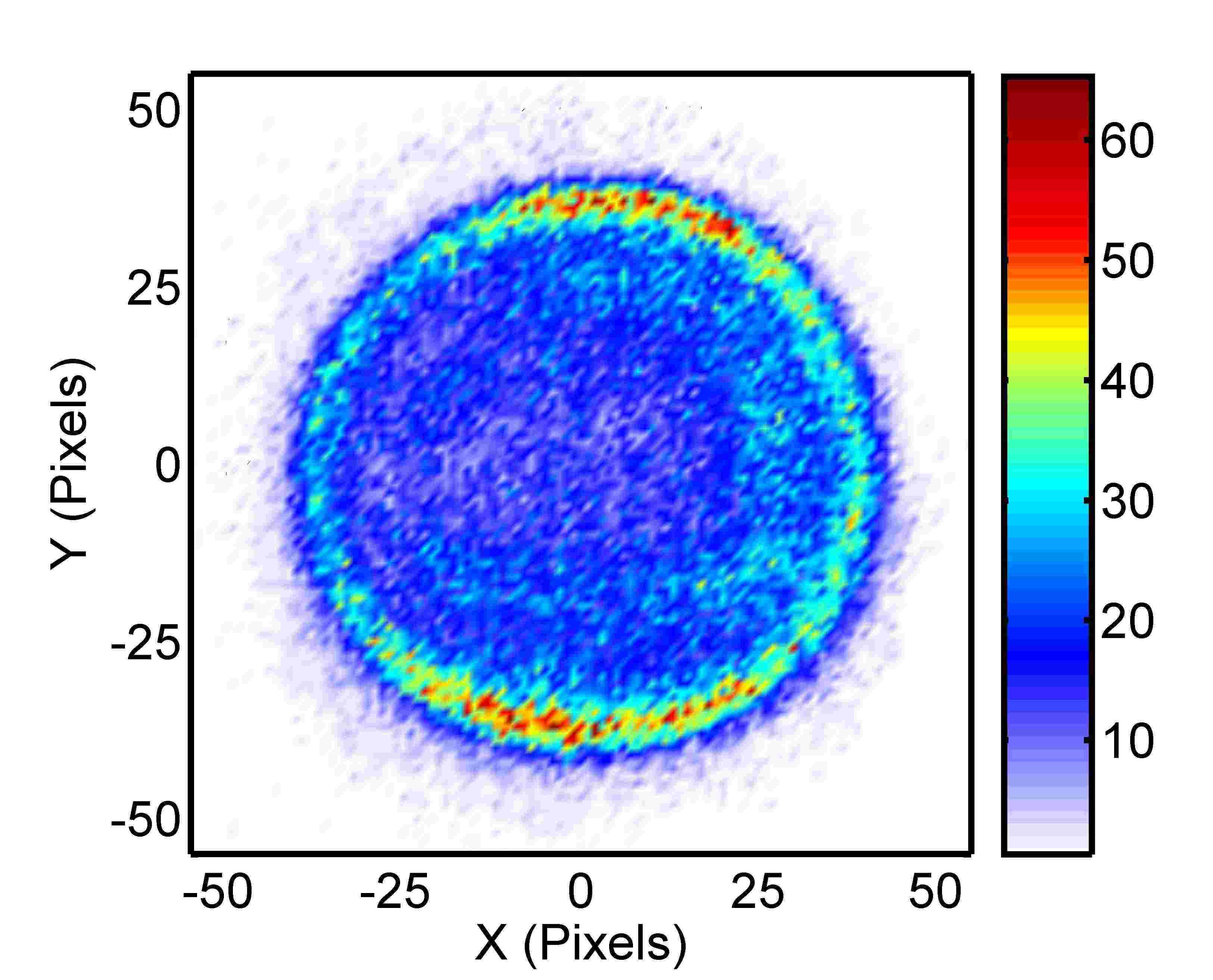}}
  \subfloat[$H^{-}$ at 10.4 eV]{\includegraphics[width=0.3\columnwidth]{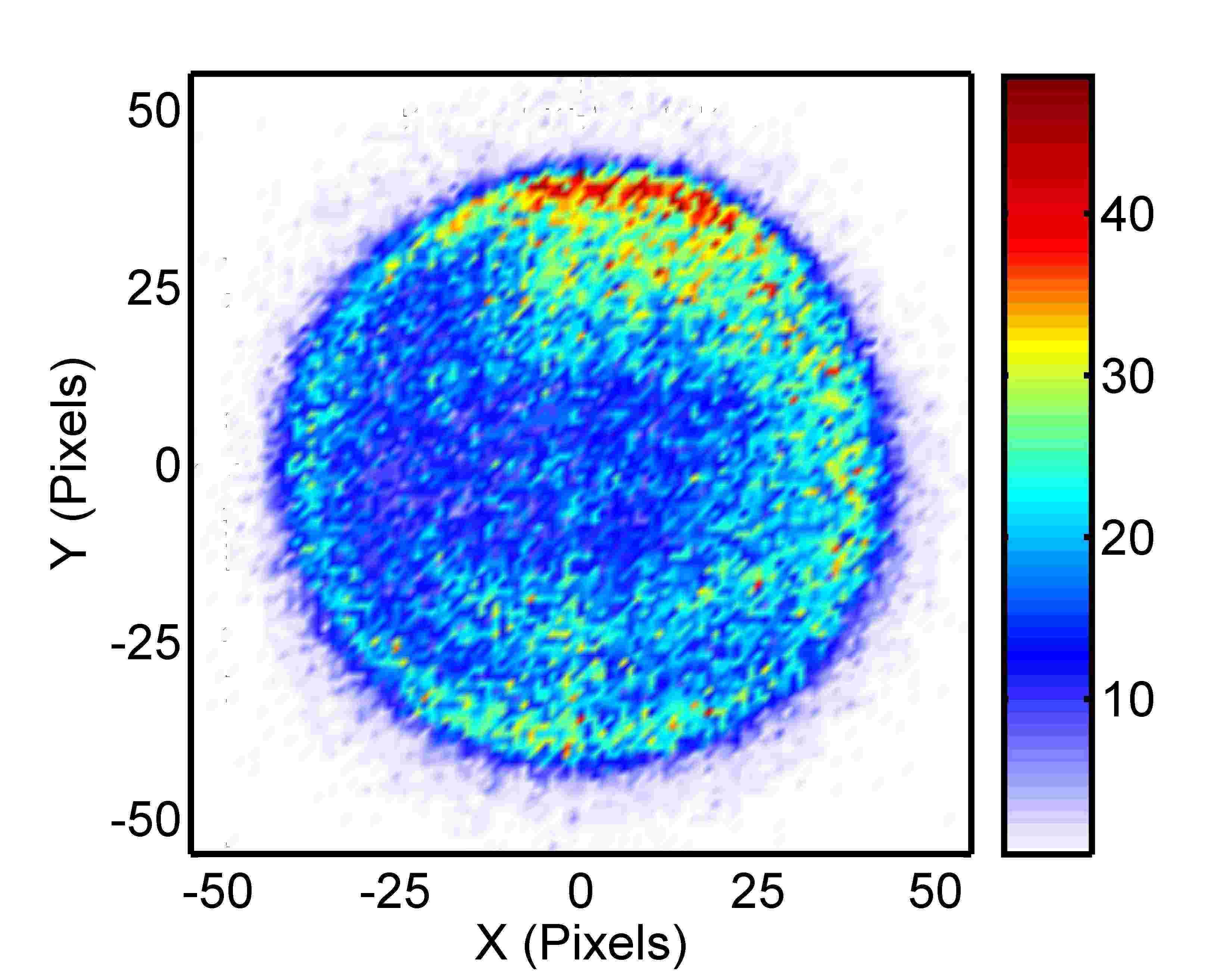}}
 \caption{Velocity images of $H^{-}$ ions from DEA to $H_{2}S$ at various electron energies. The electron beam direction is from top to bottom in every image.}
 \label{fig4.2}
 \end{figure}
 
\begin{figure}[!h]
\centering
\subfloat[$S^{-}/SH^{-}$ at 1.0 eV]{\includegraphics[width=0.25\columnwidth]{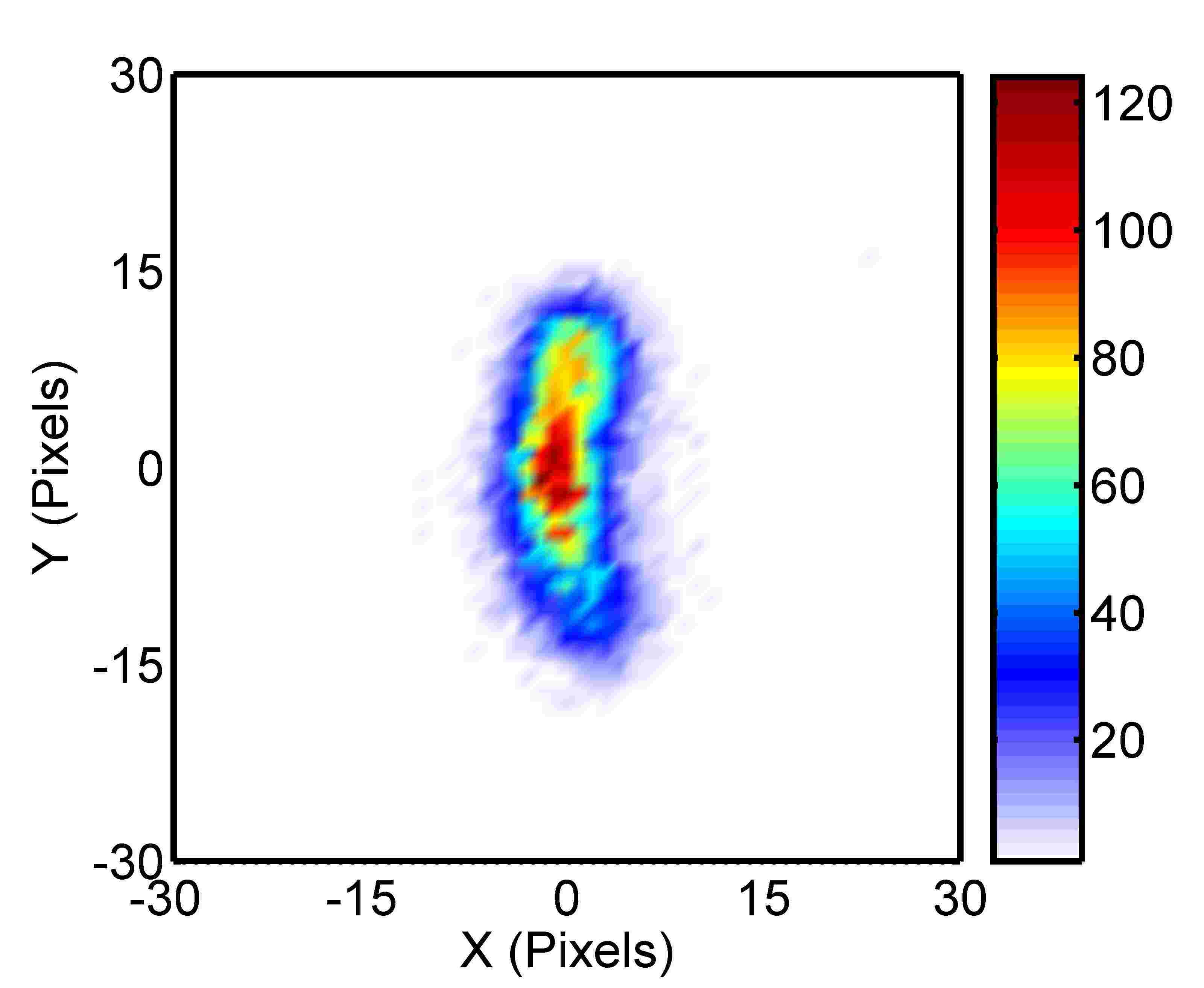}}
\subfloat[$S^{-}/SH^{-}$ at 2.0 eV]{\includegraphics[width=0.25\columnwidth]{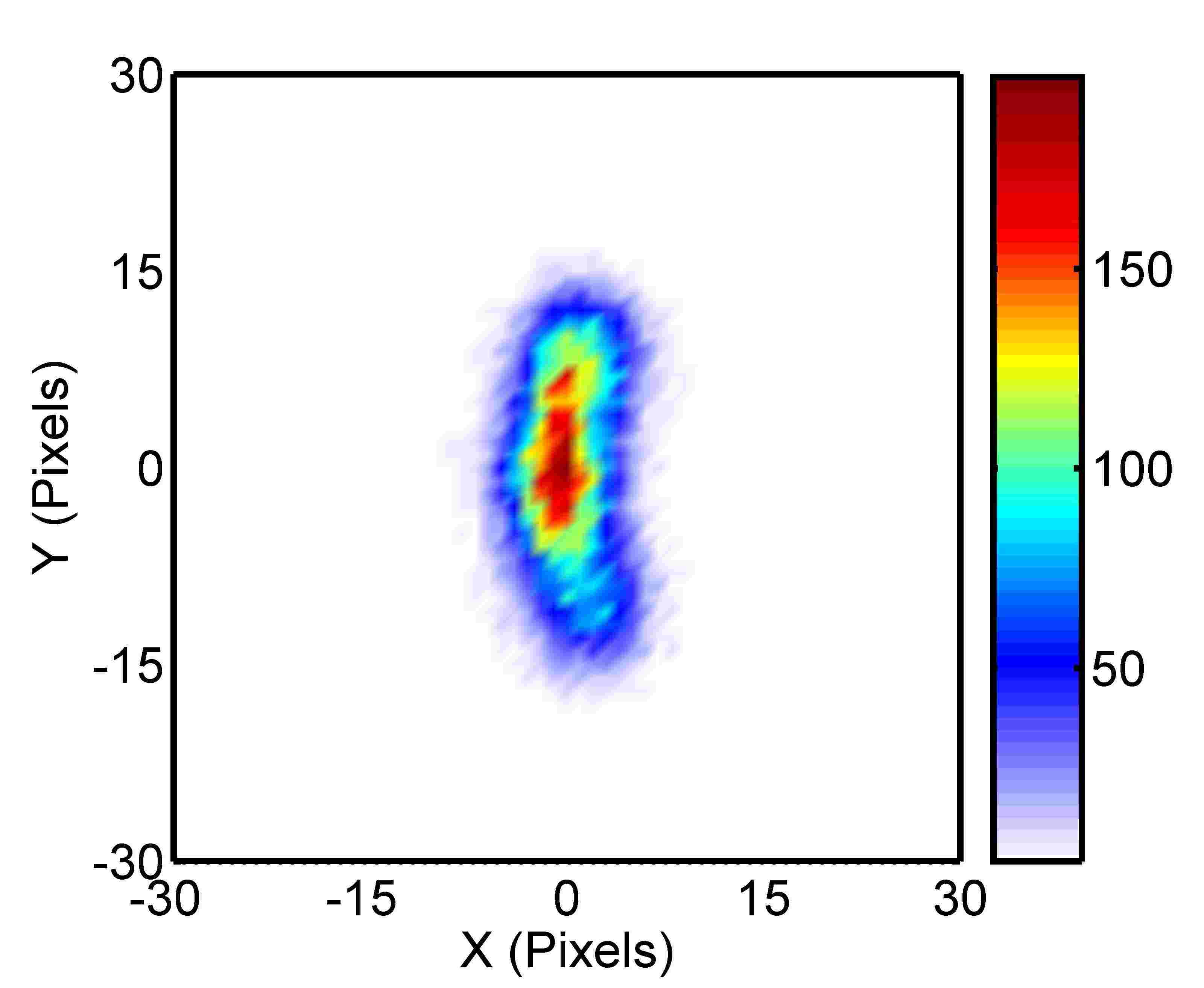}}
\subfloat[$S^{-}/SH^{-}$ at 3.0 eV]{\includegraphics[width=0.25\columnwidth]{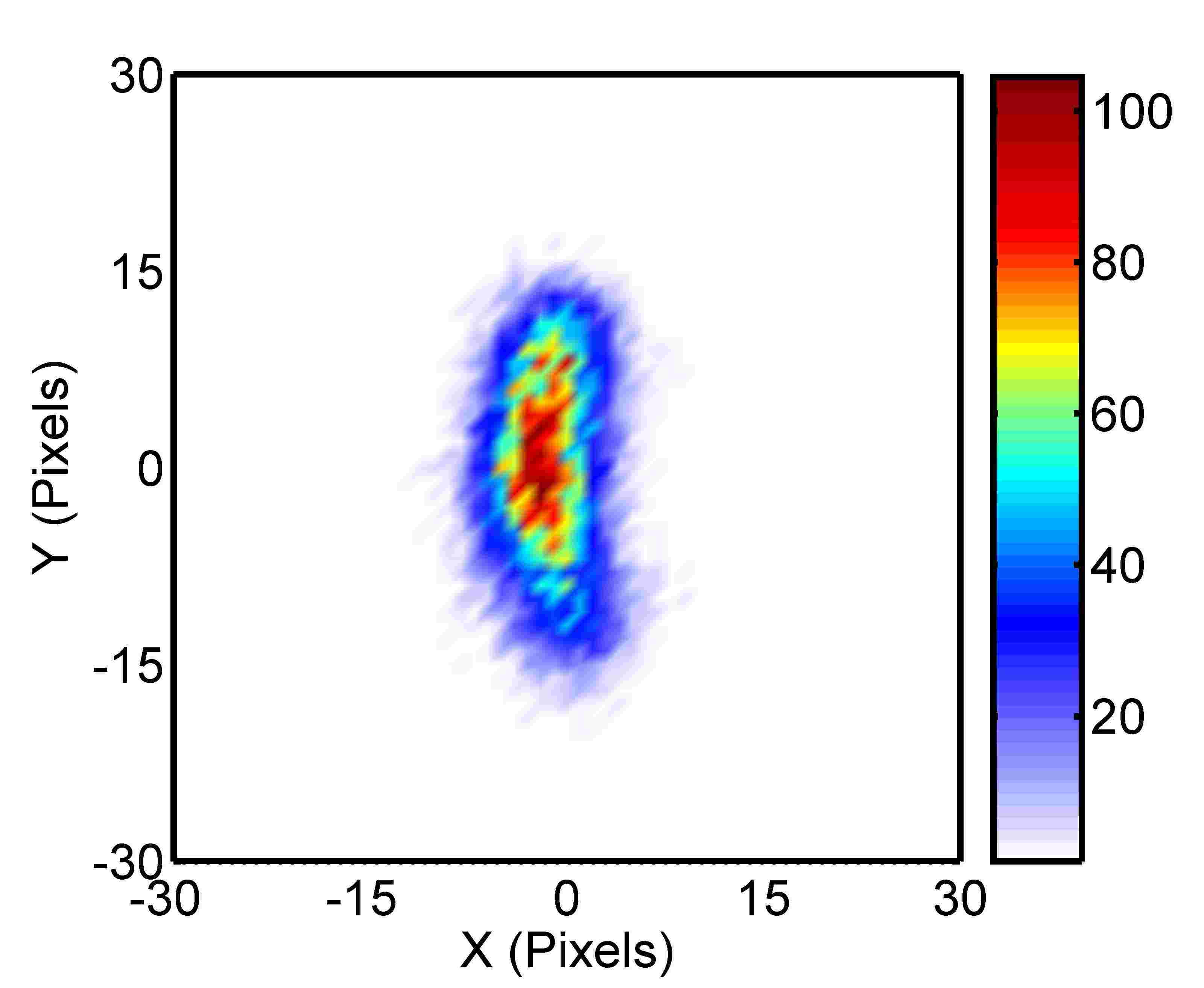}}
\subfloat[$S^{-}/SH^{-}$ at 5.5 eV]{\includegraphics[width=0.25\columnwidth]{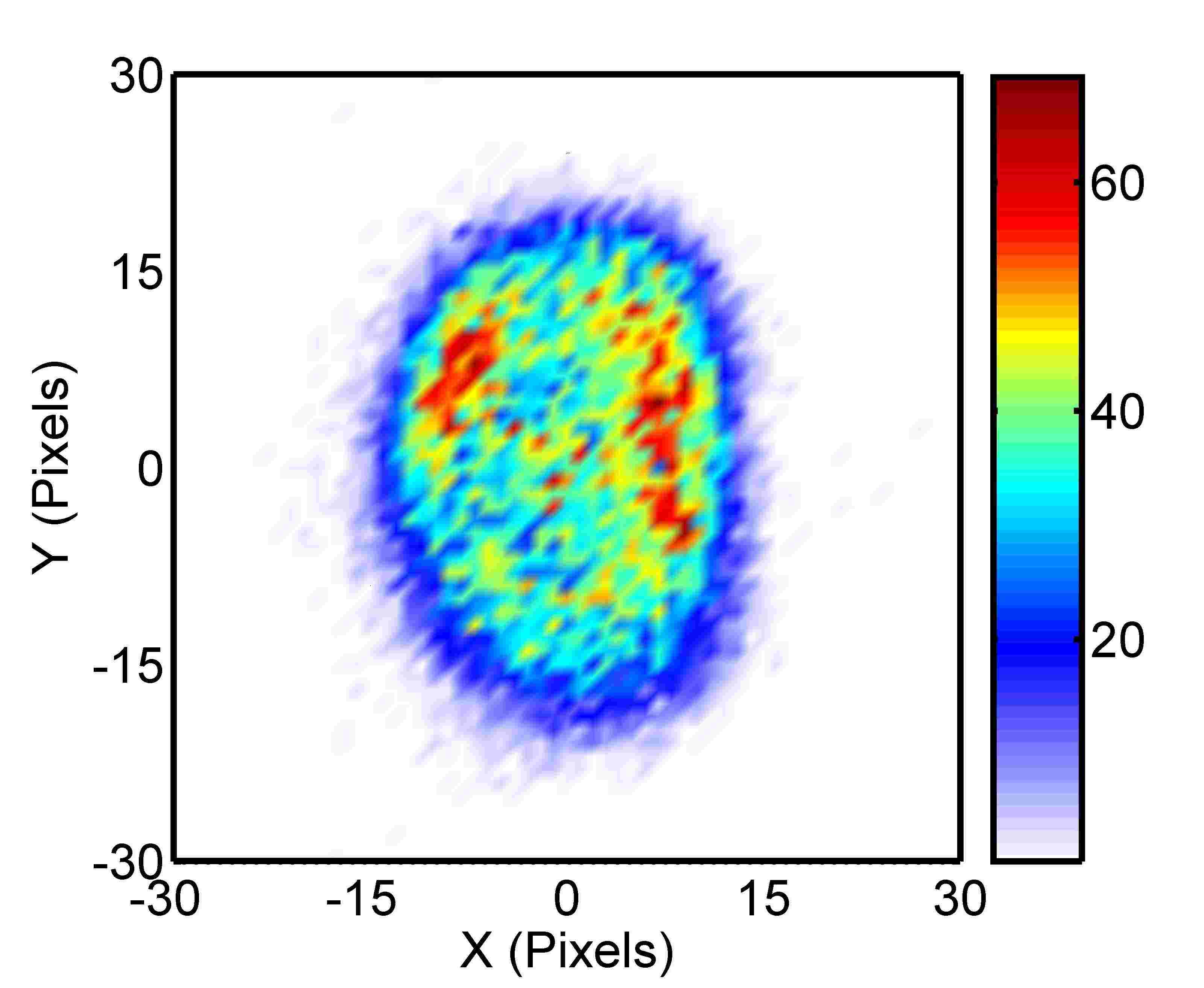}}\\
\subfloat[$S^{-}/SH^{-}$ at 7.5 eV]{\includegraphics[width=0.25\columnwidth]{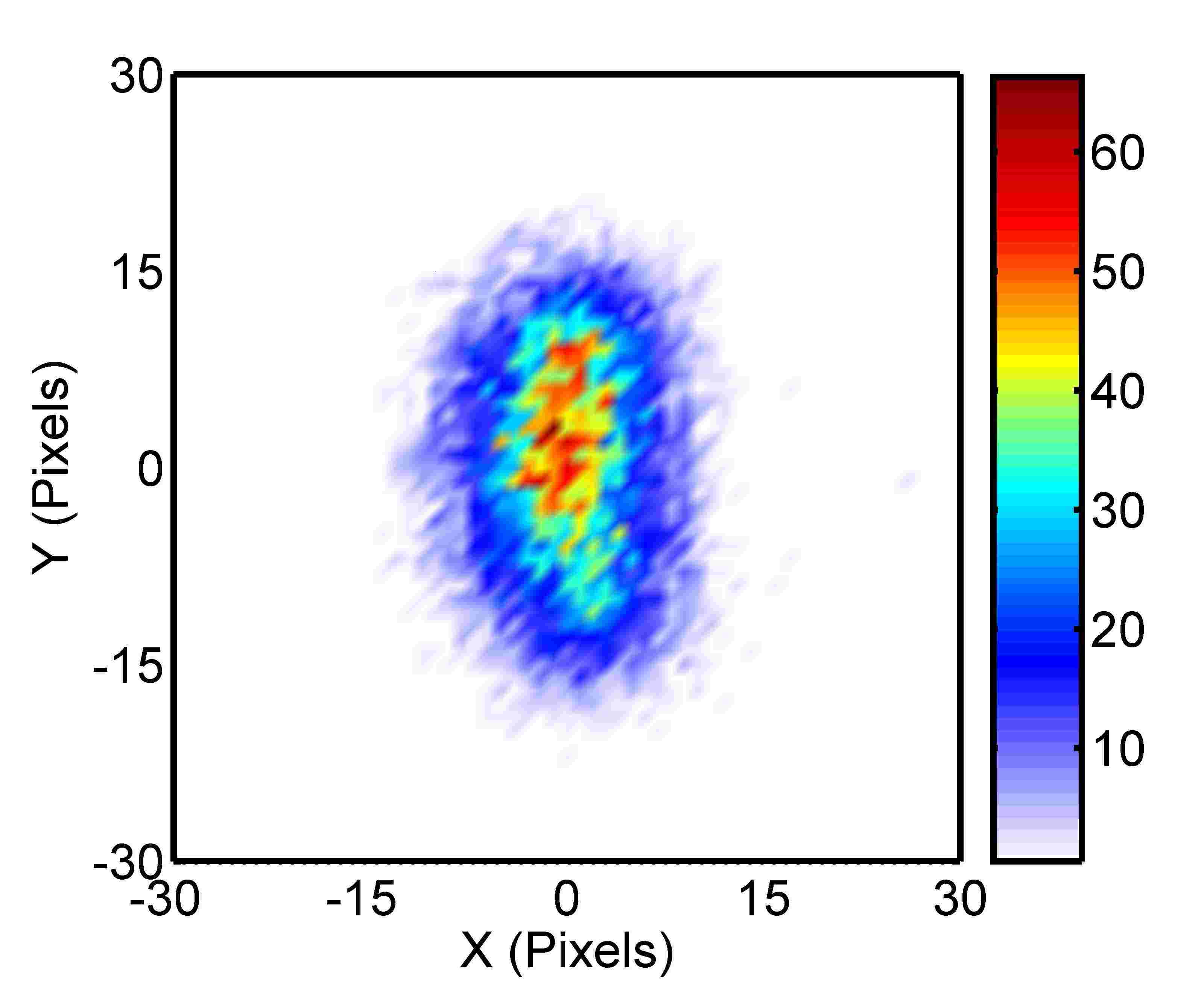}}
\subfloat[$S^{-}/SH^{-}$ at 8.5 eV]{\includegraphics[width=0.25\columnwidth]{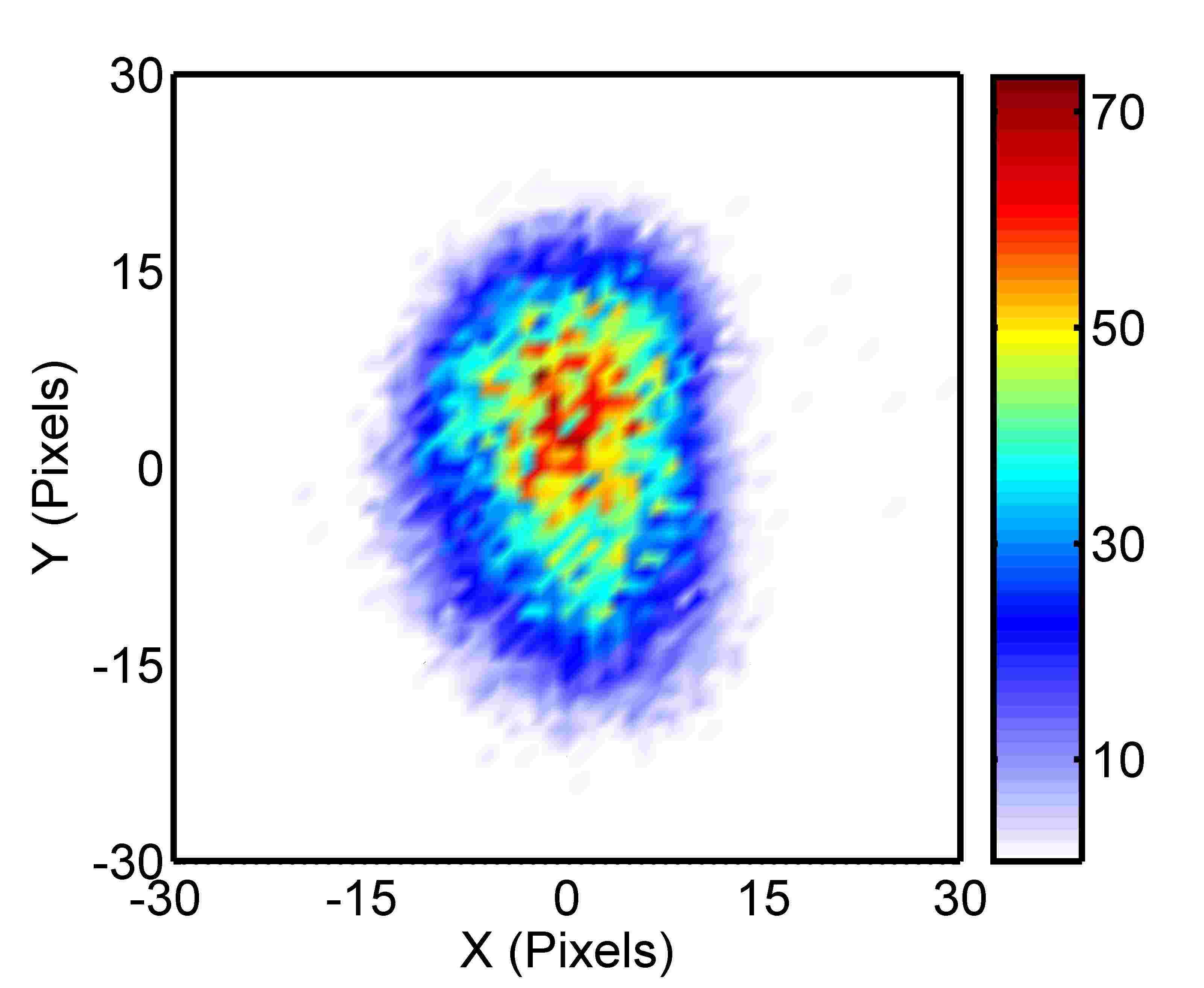}}
\subfloat[$S^{-}/SH^{-}$ at 9.6 eV]{\includegraphics[width=0.25\columnwidth]{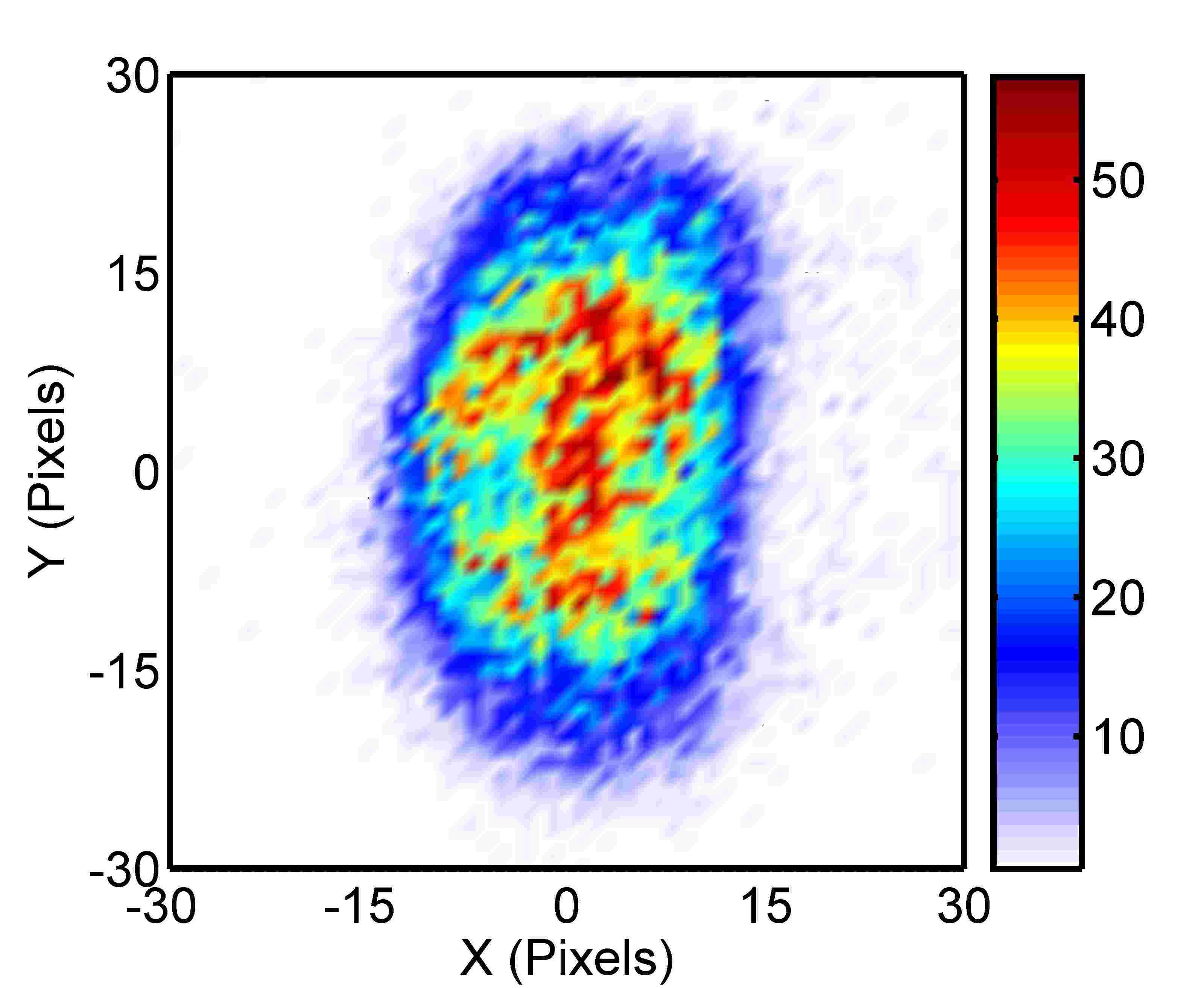}}
\subfloat[$S^{-}/SH^{-}$ at 10.4 eV]{\includegraphics[width=0.25\columnwidth]{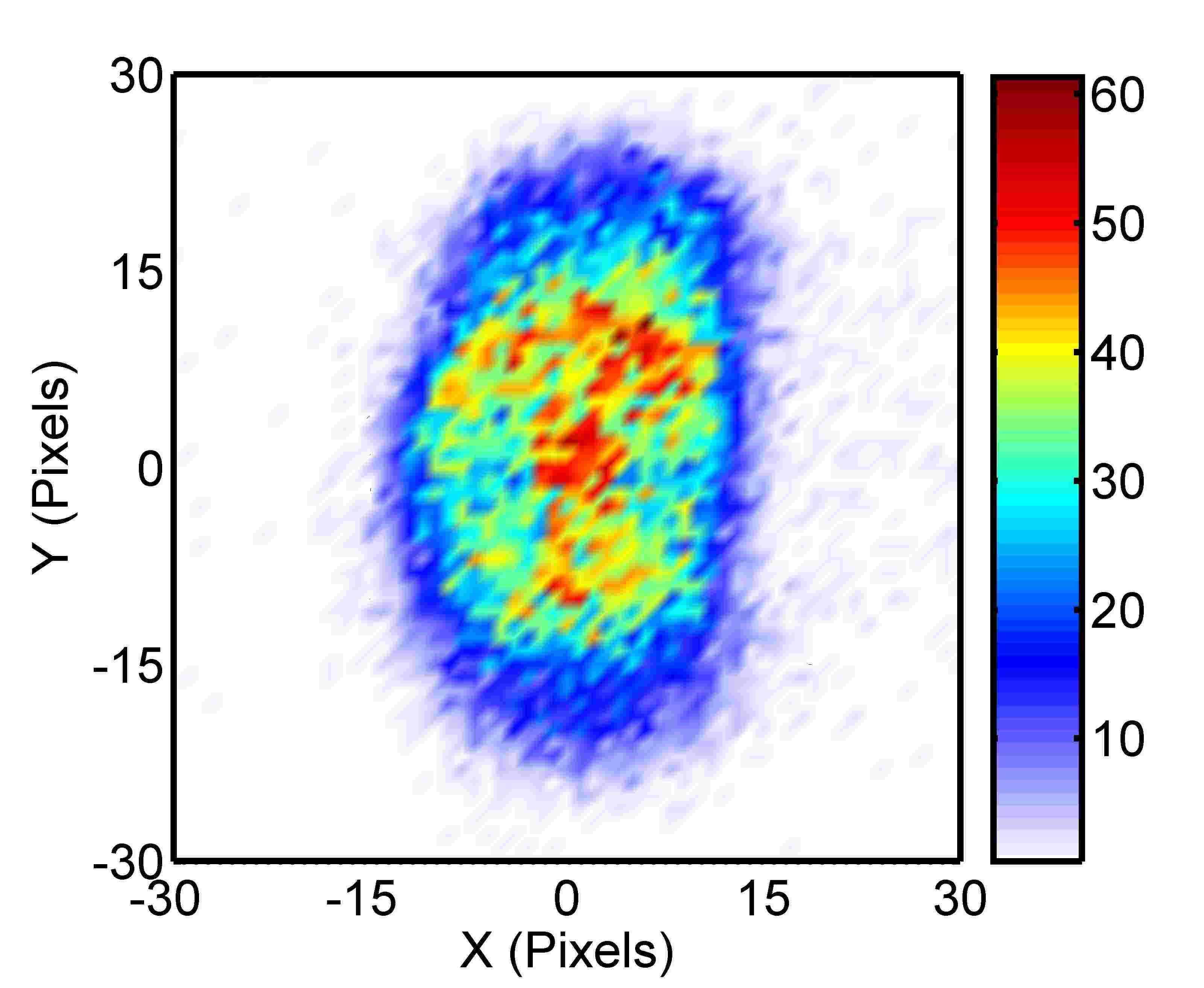}}
\caption{Velocity images of $S^{-}$/$SH^{-}$ from DEA to $H_{2}S$. The electron beam direction is from top to bottom in every image.} 
\label{fig4.3}
\end{figure}

Figure \ref{fig4.1} shows the ion yield curves of \ce{H-} and \ce{S-}/\ce{SH-}  (unresolved) fragment ions produced from DEA to \ce{H2S}. One can observe clear peaks in the \ce{H-} curve at 5.2 eV, 7.5 eV and a broad shoulder centered at 9.6 eV. \ce{S-}/\ce{SH-} ion yield has strong peaks at 2.4 and 9.6 eV with a small peak between 5 and 6 eV. The velocity images of \ce{H-} and \ce{S-}/\ce{SH-} ions are given in Figure \ref{fig4.2} and \ref{fig4.3} for the incident electron energy range 1 to 10 eV. The details of the kinetic energy and angular distribution of the fragment ions produced at 5.2 eV, 7.5 eV and 9.6 eV resonances are discussed below. Since \ce{H2S} belongs to \ce{C_{2v}} point group, the angular distribution expressions are same as those derived in the preceding paper on water and are used to fit the angular distribution data in the case of \ce{H2S}. The only parameter affected is $\beta$ i.e. the angle between dissociation axis and molecular symmetry axis (\ce{C2}). As H-S-H bond angle is $95^{\circ}$ in ground state equilibrium, $\beta$ is $47.5^{\circ}$ for \ce{H-} produced from dissociation of a SH bond in ground state equilibrium geometry of neutral \ce{H2S}.

\subsection{First resonance process at 2.4 eV}

\subsubsection*{\ce{S-} and \ce{SH-} ions}

From Figure \ref{fig4.1}, we see that \ce{S_}/\ce{SH-} are the dominant ion fragments and start appearing at energies below 1.0 eV. Figures \ref{fig4.3} (a), (b) and (c) show the velocity images of \ce{S_}/\ce{SH-} ions at electron energies 1eV, 2 eV and 3 eV respectively, seen as a central blob with finite intensities in the forward and backward directions giving a vertically elongated structure. This elongation, we believe, is due to the extended interaction volume along the incident electron beam with the effusive gas beam and imaging distortions. The kinetic energies estimated for these ions from thermodynamic thresholds are about few tens of milli electron Volts. The recent work by Abouaf and Billy \cite{c4abouaf} on \ce{S-} and \ce{SH-} fragment yields from DEA to \ce{H2S} in the electron energy range 0-4 eV with improved electron beam resolution of 0.040 eV gives a lot of information on the potential energy surface leading to the formation of \ce{S-} and \ce{SH-}. The \ce{SH-} ion yield showed a vertical onset at 1.6 eV (thermodynamic threshold is 1.58 eV) suggestive of an attractive potential energy surface. Whereas, \ce{S-} ion yield curve starts at 0.6 eV and has no vertical onset. The ion yield curves of both ions show structures characteristic of vibrational energy distribution in their ion yield curves indicating symmetric or anti-symmetric vibrational modes of the neutral \ce{H2S} molecule. In comparison, \ce{O-} and \ce{OH-} appearance thresholds are 3.57 and 3.29 eV respectively in water, but there is no resonance below 6.5 eV where they appear. Azria et al. \cite{c4azria2} attribute this absence of a shape resonance in water due to negligible Franck Condon overlap and low survival probability.

\subsection{Second resonance process at 5.2 eV}

\subsubsection{\ce{H-} ions}
\begin{figure}[!ht]
\centering
\subfloat[]{\includegraphics[width=0.33\columnwidth]{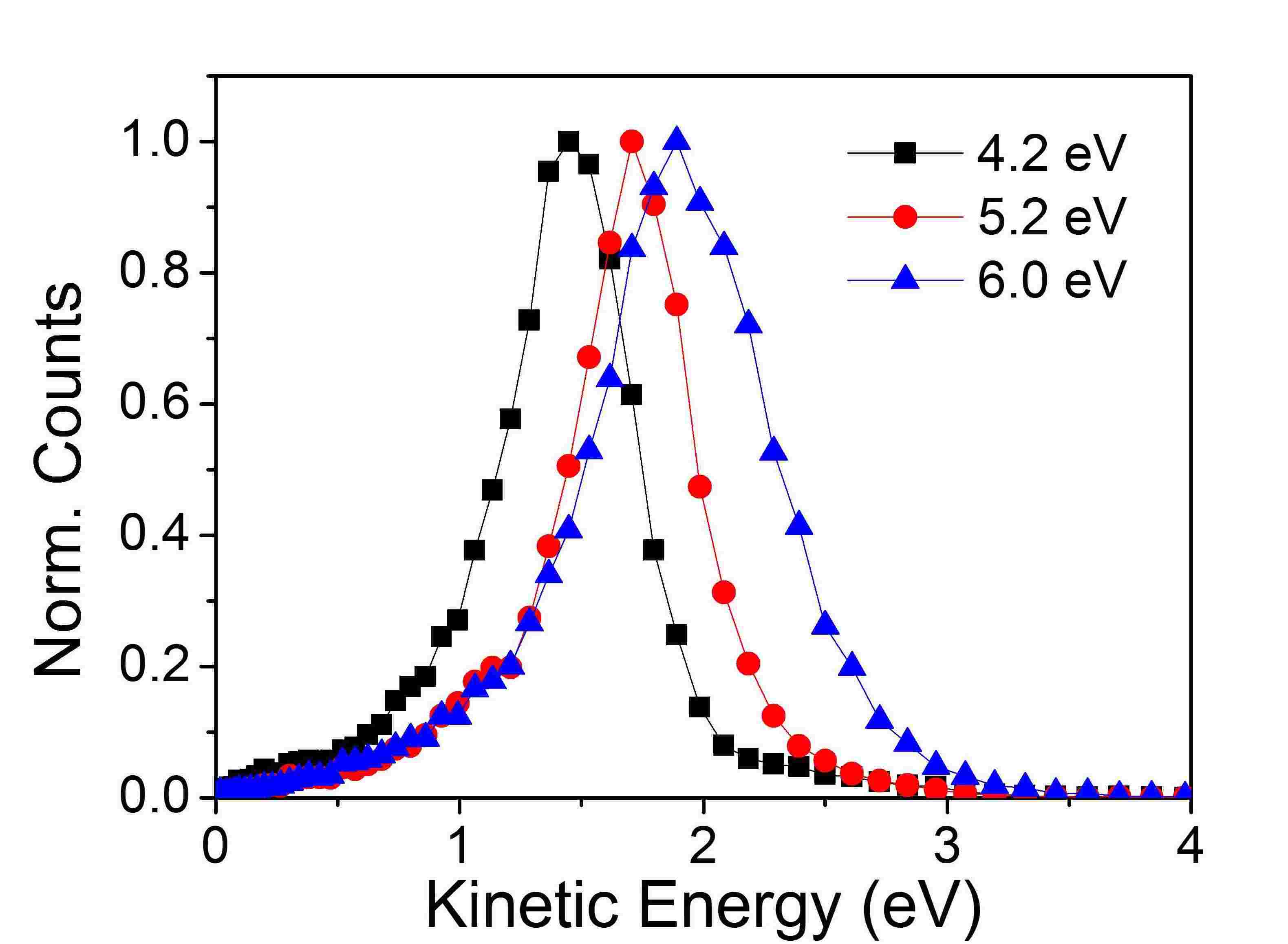}}
\subfloat[]{\includegraphics[width=0.33\columnwidth]{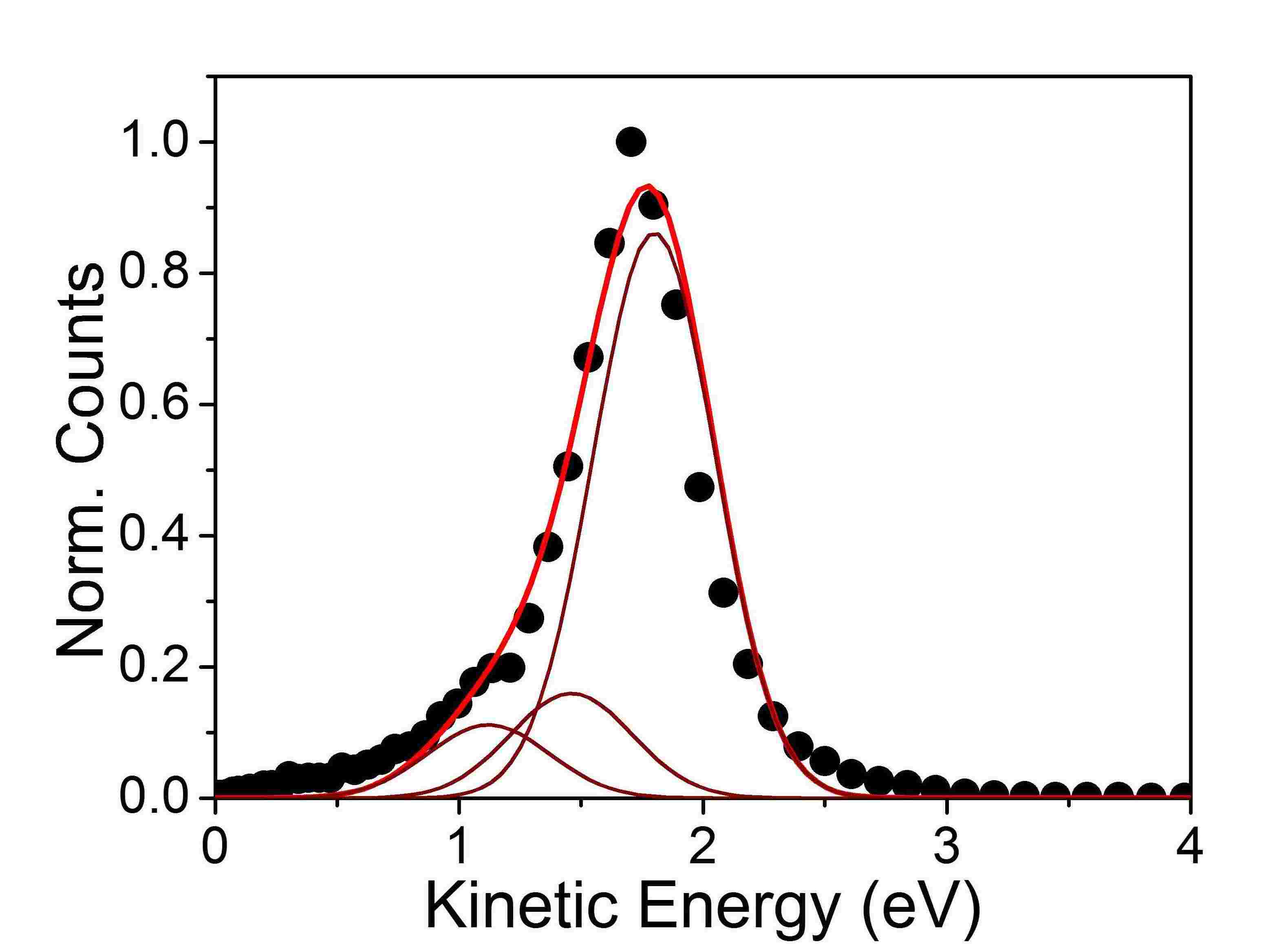}}
\subfloat[]{\includegraphics[width=0.33\columnwidth]{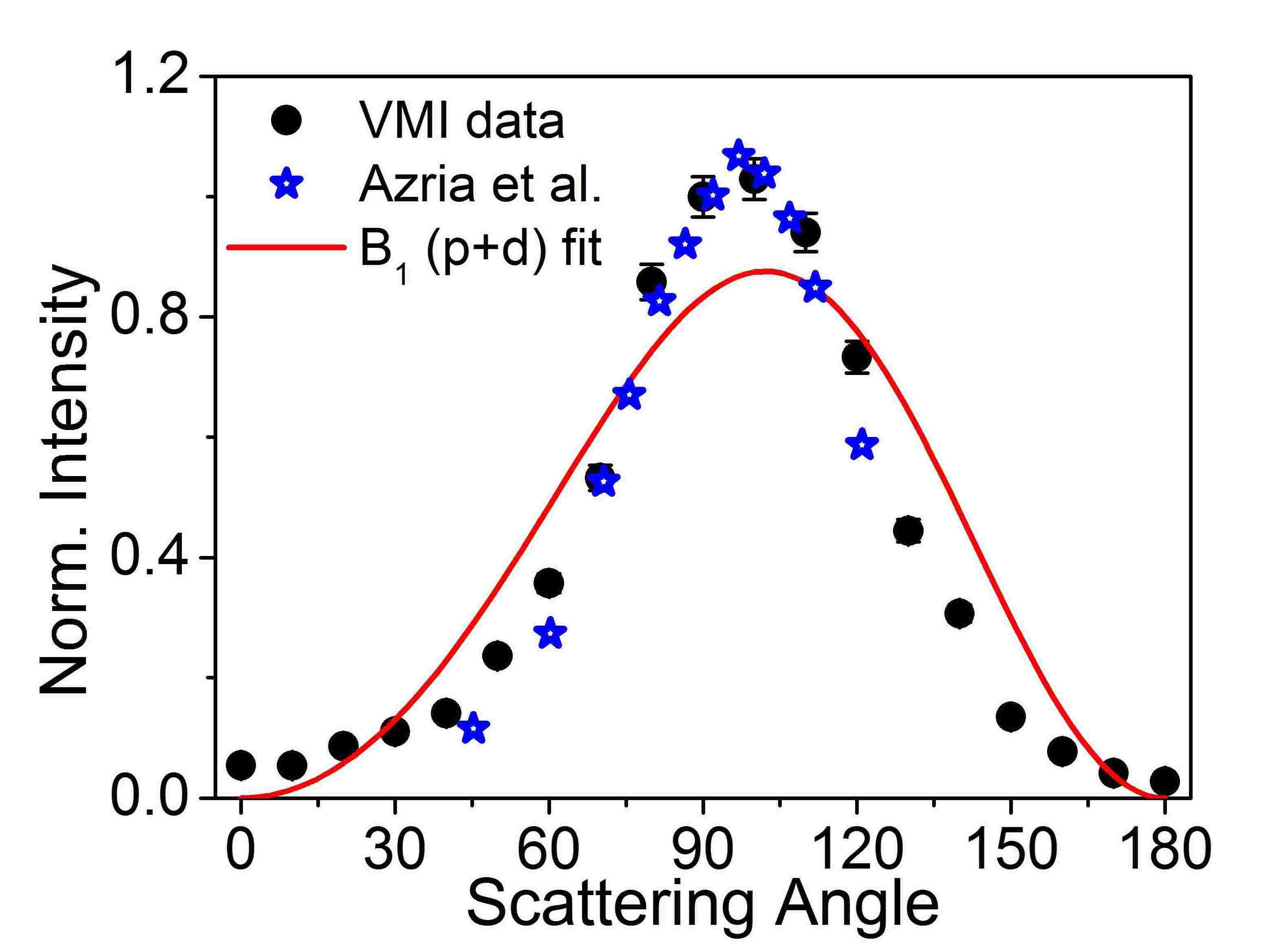}}
\caption{(a) Kinetic energy distribution of \ce{H-} ions across the second resonance process at 4.2 eV, 5.2 eV and 6.0 eV. (b) KE distribution of \ce{H-}  at 5.2 eV fitted with Gaussian functions to determine the intensity of vibrational excitation. (c) Angular distribution of \ce{H-} ions of all kinetic energies at 5.2 eV compared with the data of Azria et al. \cite{c4azria} taken at 5.35 eV. The angular distribution is fit with \ce{B1} symmetry functions involving $p$ and $d$ partial waves. The best fit is given for p:d = 1:0.2 with a phase difference of $\pi$ radians.}
\label{fig4.4}
\end{figure}

The second resonance peaking at 5.2 eV produces \ce{H-} ions dominantly. The velocity images of the \ce{H-} arising from this resonance process  are shown in Figure \ref{fig4.2}(a), (b) and (c) for electron energies 4.2 eV, 5.2 eV and 6 eV respectively. The kinetic energy distribution of \ce{H-} ions across the resonance is shown in Figure \ref{fig4.4}(a). The maximum kinetic energy is seen to be about 2.3 eV at 5.2 eV incident energy and suggests the dissociation channel to be \ce{H-} + SH (X $^{2}\Pi)$ (threshold energy 3.15 eV) where the neutral \ce{SH} fragment is in electron ground state but vibrationally excited. To elaborate, the maximum kinetic energy of \ce{H-} in this channel would be 33/34th of (5.2 - 3.15) eV i.e. approximately 2 eV. And we observe a value of 2.3 eV close to the estimated value. The peaks in the kinetic energy distribution shift towards higher values with marginal increase in the width. This indicates that the excess energy is channelled into translational energy of the \ce{H-} and \ce{SH_{$\nu$=0}} fragments with little energy going into the excitation of higher vibrational modes.  Internal excitation of the SH fragment in the form of vibrations and rotations results in broadening of the width of the distribution. We tried to determine the relative intensities of the SH vibrational states by fitting with Gaussian functions (same width, different height) - as mentioned in the previous chapter for water - representing the spread of \ce{H-} kinetic energy at positions separated by the vibrational spacing i.e. 2615 $cm^{-1}$ or 0.34 eV \cite{c4nist,c4shimanouchi,c4herzberg2}. The fit is shown in Figure \ref{fig4.4}(b). This isn't a very accurate way of determining the vibrational state intensities but gives a fairly good idea of the number of vibrational states populated. We find that ground vibrational state is maximally populated whereas $\nu$=1 and $\nu$=2 are also populated meagrely. The relative intensities of the states found from the best fit of the Gaussian functions are approximately in the ratio 1:0.2:0.1. Similar exercise in the case of 6.5 eV resonance in water shows vibrational states populated up to v=4 and the relative intensities are 1:0.82:0.57:0.32:0.25. This shows that most of the excess energy is converted into translational energy rather than internal excitation of the fragments in the case of \ce{H2S}. At 6 eV, the KE distribution is broader and suggests significant population of $\nu$=1 and $\nu$=2 levels.

The angular distribution of $H^{-}$ ions is shown in Figure \ref{fig4.4}(c) and we see a distribution peaking close to $100^{\circ}$ with a slight asymmetry about the peak. The distribution is very similar to that reported by Azria et al. \cite{c4azria} as shown in the Figure \ref{fig4.4}(c).  This distribution indicates \ce{B1} symmetry of the \ce{H2S^{-*}} anion caused by the \ce{2b1 -> 4sa1} excitation.This  distribution indicates \ce{B1} symmetry of the \ce{H2S-^{*}} anion caused by the \ce{2b1 -> 4sa1} excitation. Fitting the angular distribution using \ce{B1} symmetry functions shows $p$ and $d$ partial waves involved in the scattering process. The asymmetry about the $90^{\circ}$ direction is due to mixing of the $p$ and $d$ wave with a phase difference of $\pi$ radians as can be seen from the fit in Figure \ref{fig4.4}(c).  These observations and fits are identical to the 6.5 eV resonance in water. UV absorption studies on \ce{H2S} \cite{c4robbins} show an extended absorption band/structure in the 40000 - 60000 $cm^{-1}$ (i.e. 4.9 to 7.5 ev) caused the \ce{2b1 -> 6a1} and \ce{2b1 -> 3b2} valence excitations along with Rydberg excitations to \ce{4sa1}. While the excitation to \ce{6a1} is dipole allowed, the \ce{3b2} is forbidden optically. Therefore, the parent state of the DEA resonance at 5.2 eV is \ce{^{3}B1} arising from the \ce{2b1 -> 6a1} excitation consistent with the kinetic energy and angular distributions and symmetry rules.

\begin{figure}[!ht]
\centering
\subfloat[]{\includegraphics[width=0.5\columnwidth]{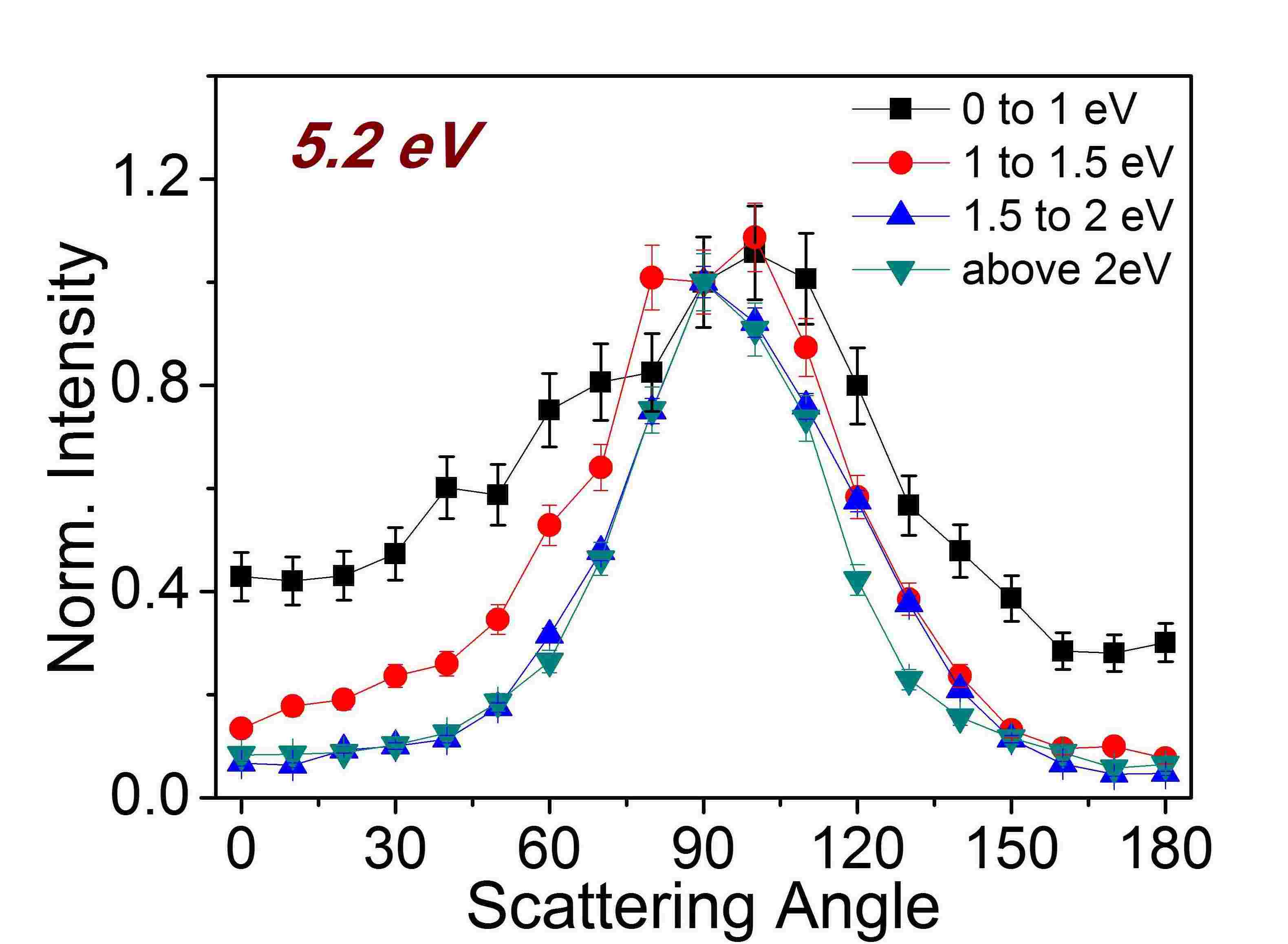}}
\subfloat[]{\includegraphics[width=0.5\columnwidth]{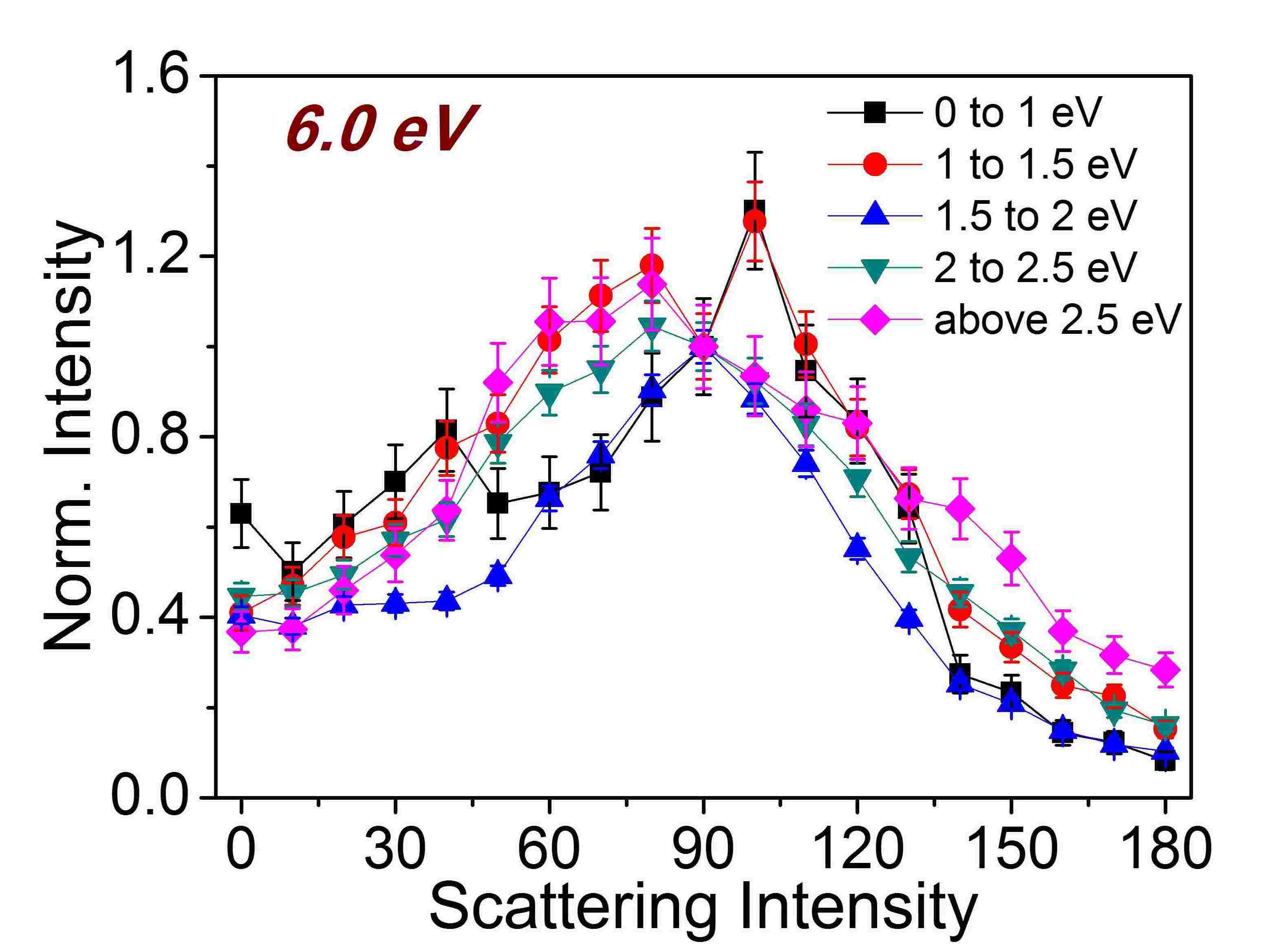}}
\caption{Angular distribution of \ce{H-} ions at 5.2 eV and 6.0 eV shown as a function of the kinetic energy.}
\label{fig4.5}
\end{figure}

Azria et al. \cite{c4azria} reported that at a higher electron energy (at 5.97 eV), the angular distribution of \ce{H-} ions corresponding to the $\nu$=0 and $\nu$=1 states of the SH fragment are different. This is not seen for \ce{H-} ions from \ce{H2O}. Haxton et al. \cite{c4h4} showed in their calculations that taking different entrance amplitude for each vibration state of SH could lead to difference in the angular distribution of \ce{H-} ions. To verify this, we looked at the angular distributions of \ce{H-} ions at 5.2 eV and 6.0 eV as a function of the kinetic energy (see Figure \ref{fig4.5}). At 6.0 eV, forward scattering angles are more intense than at 5.2 eV and we see that the angular distributions are same for the \ce{H-} ions with KE above 2.5 eV (corresponding to \ce{SH_{$\nu$=0}}) and for 2-2.5 eV (for \ce{SH_{$\nu$=1,2}}). Our kinetic energy resolution is not sufficient to clearly separate the $\nu=0$ and $\nu=1$ states of SH and hence, cannot conclusively confirm angular distribution of \ce{H-} ions being different for the two vibrational states. While, it may be noted that the forward scattering of the \ce{H-} ions is seen at the 7.5 eV resonance, Azria et al. \cite{c4azria} have ruled out the contribution from the tail of the second process stating that the \ce{H-} ions from the \ce{A1} resonance are mostly associated with SH fragment in $\nu=0$ state. However, in our measurements, as seen from the plots in Figure \ref{fig4.5}(b), the contribution from the 7.5 eV resonance seems to cause the enhanced forward scattering of \ce{H-} at all kinetic energies.

\subsubsection*{\ce{S-} ions}

\begin{figure}[!htbp]
\centering
\includegraphics[width=0.5\columnwidth]{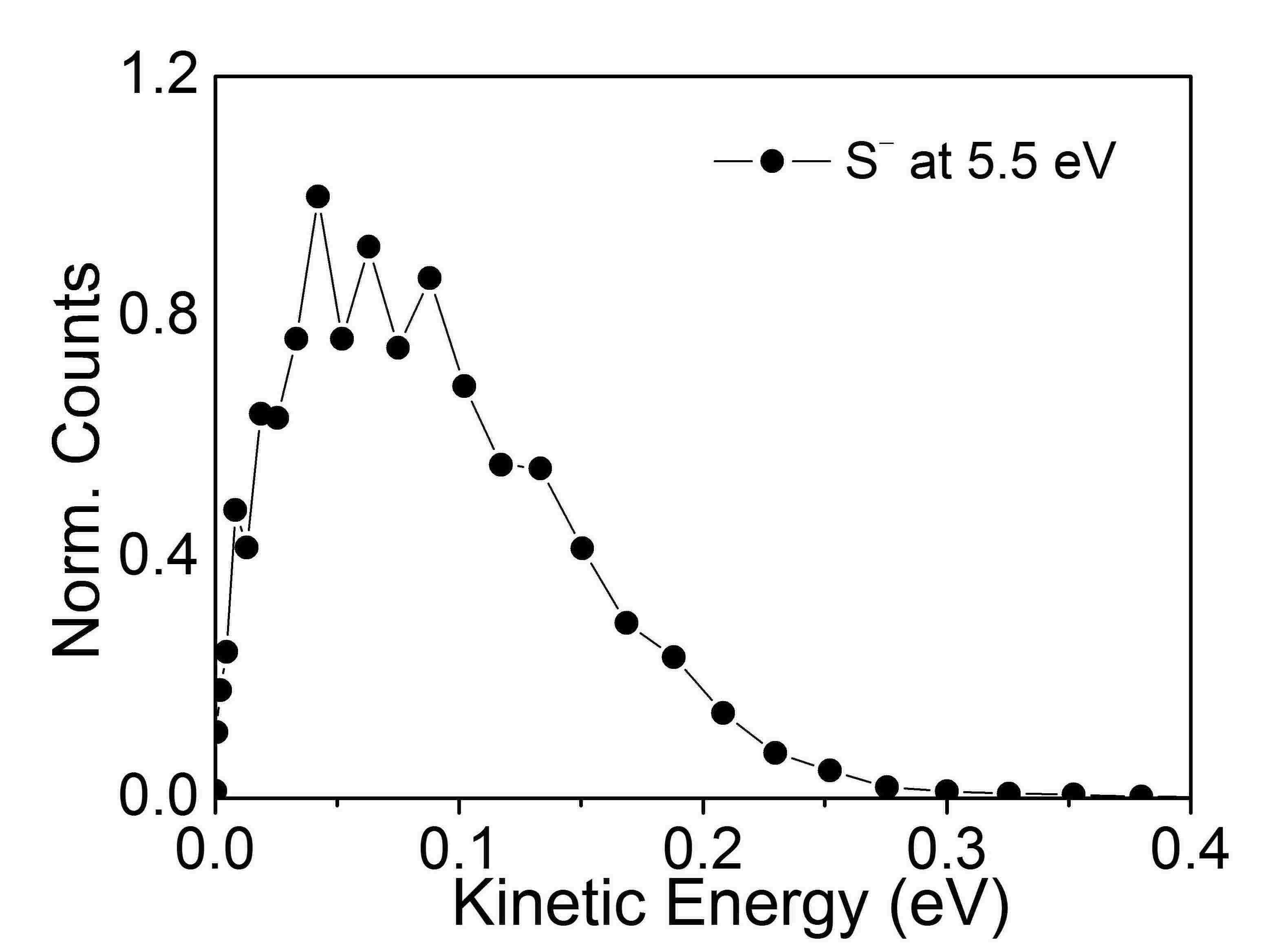}
\caption{Kinetic energy distribution of $S^{-}$ ions at 5.5 eV}
\label{fig4.6}
\end{figure}

Figure \ref{fig4.3}(d) shows the velocity image of \ce{S-}/\ce{SH-} (unresolved) ions produced at this resonance. The higher resolution measurements of the mass spectrum and ion yield curves by Abouaf and Teillet-Billy \cite{c4abouaf} found no \ce{SH-}  ions at this resonance. Although the presence of \ce{SH-}$(^{1}\Sigma^{+})$ + H channel is energetically possible, it is ruled out based on Wigner-Witmer correlation rules, for a resonance of \ce{B1} symmetry.  Figure \ref{fig4.3}(d) show some anisotropic distribution of the ions scattered perpendicular to the electron beam indicating that these might be \ce{S-} ions ejected via the \ce{S-} + \ce{H2} channel. We see that the image shows a left-right asymmetry due to imaging distortion. However, the perpendicular scattering is discernible. This is on expected lines from a resonance of \ce{B1} symmetry where the \ce{H2S} molecular plane is perpendicular to electron beam. Thus, \ce{S-} ejected in the molecular plane will appear perpendicular with respect to the electron beam. The simplest mechanism for the formation of \ce{S-} is through the three-body fragmentation channel involving breaking of the two S-H bonds. However, the threshold energy for this process is 5.4 eV. We see that \ce{S-} is formed at lower energies. This can happen only if the two H atoms form a new bond between them forming \ce{H2} while breaking away from the S atom. The threshold for the \ce{S-} + \ce{H2} channel is 1.06 eV and for incident electron energies of 5 eV and 6 eV, the maximum kinetic energy of \ce{S-} estimated is 2/34th of the excess energy i.e. 0.23 eV and 0.29 eV, respectively. The KE distribution of \ce{S-} at 5.5 (Figure \ref{fig4.6}) shows maximum KE of about 0.25 eV with a peak close to 0.1 eV. The peak energy of 0.1 eV corresponds to 2.7 eV of energy in the internal excitation of \ce{H2}. This suggests that the formation of \ce{S-} is coupled with the production of \ce{H2} in highly vibrational excited states. The population of high vibrational levels in \ce{H2} is only to be expected since the internuclear separation between the two H atoms are likely to be much larger than that corresponding to $\nu$=0 state during its formation.  We also infer the presence of the three body breakup channel \ce{S- + H + H} (threshold : 5.4 eV) from the finite intensity at kinetic energies close to zero (approx. 0.03 eV).

\subsection{Third resonance process at 7.5 eV}

\subsubsection*{\ce{H-} ions}

\begin{figure}[!ht]
\centering
\subfloat[]{\includegraphics[width=0.33\columnwidth]{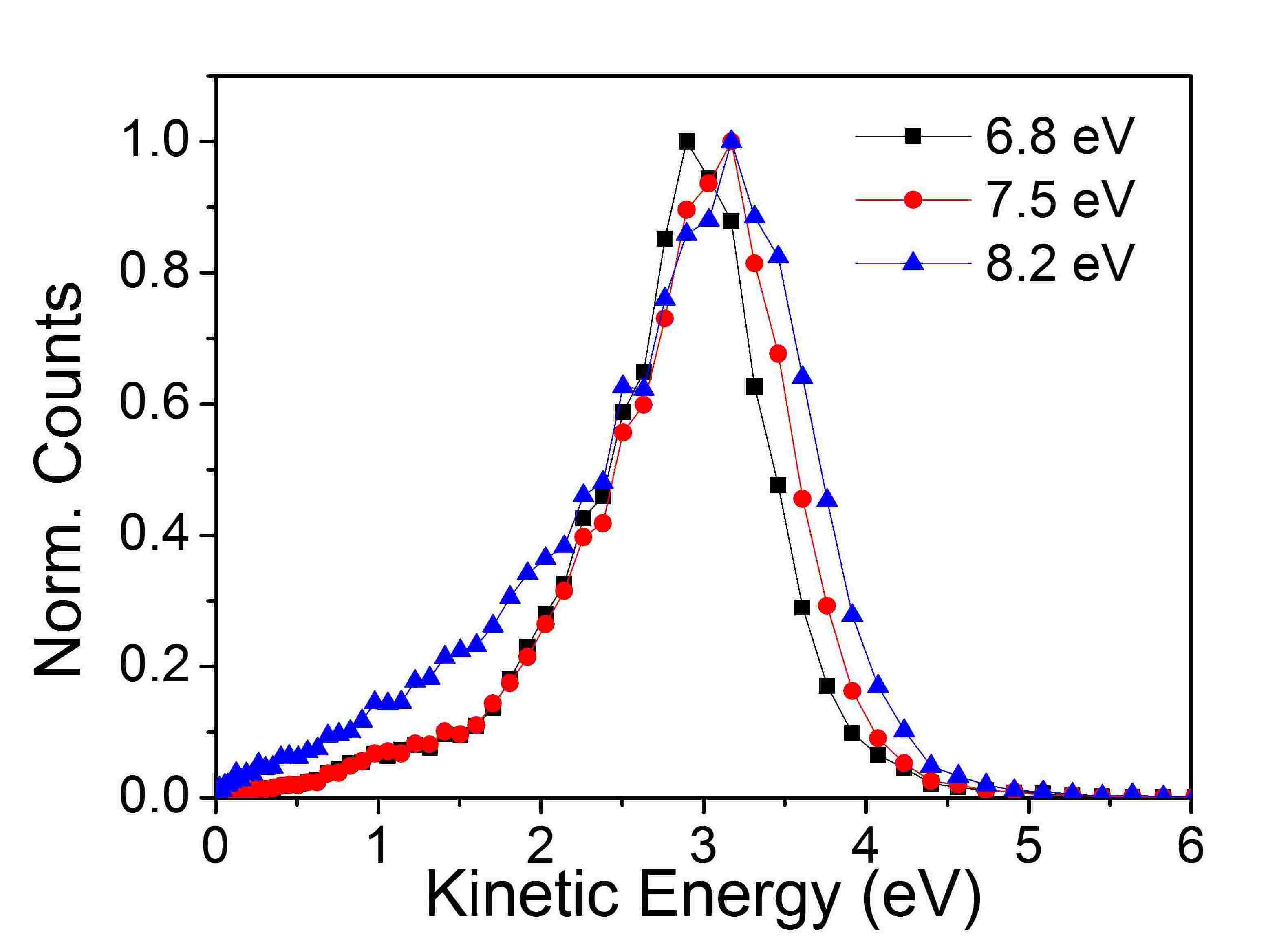}}
\subfloat[]{\includegraphics[width=0.33\columnwidth]{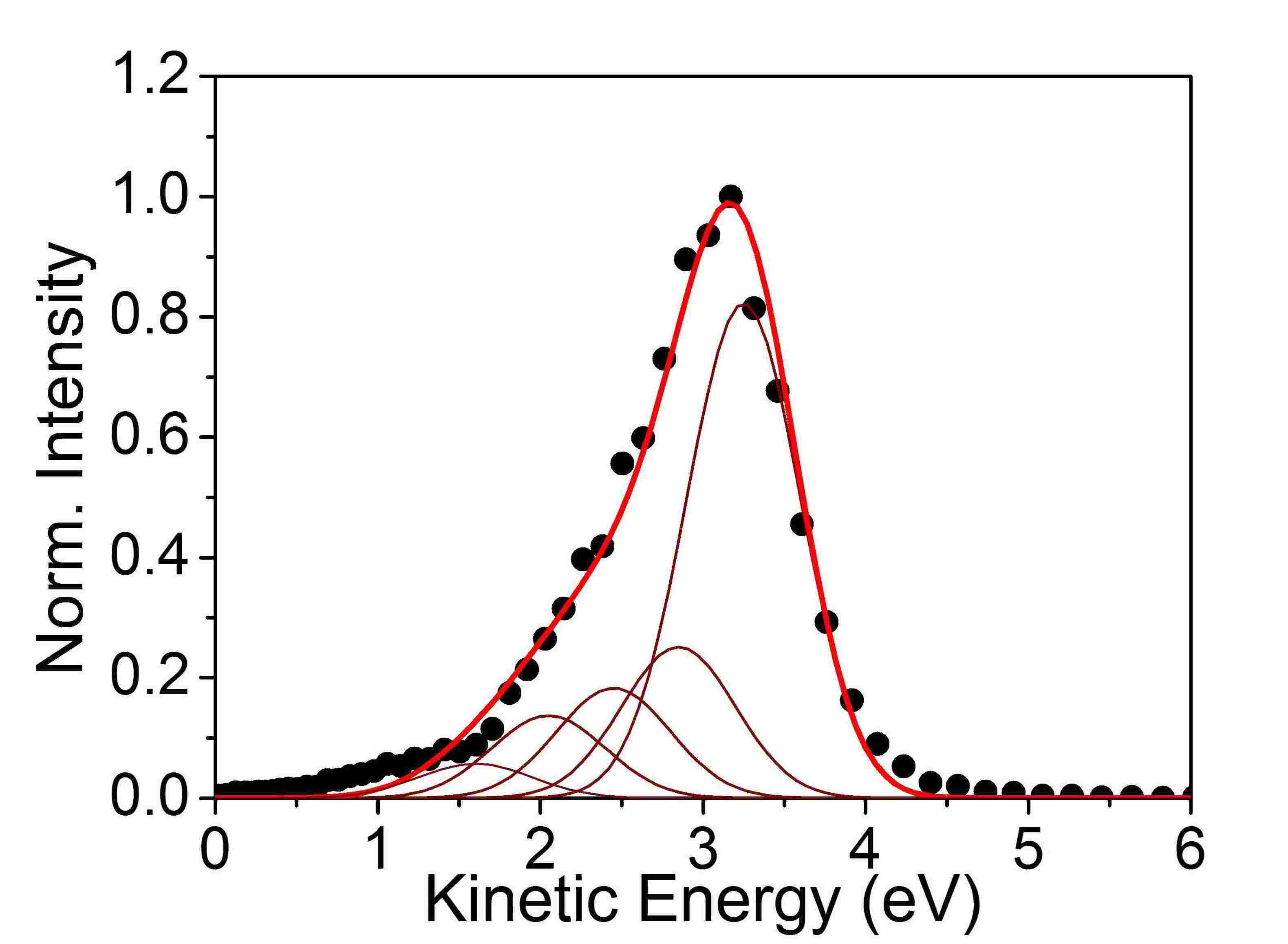}}
\subfloat[]{\includegraphics[width=0.33\columnwidth]{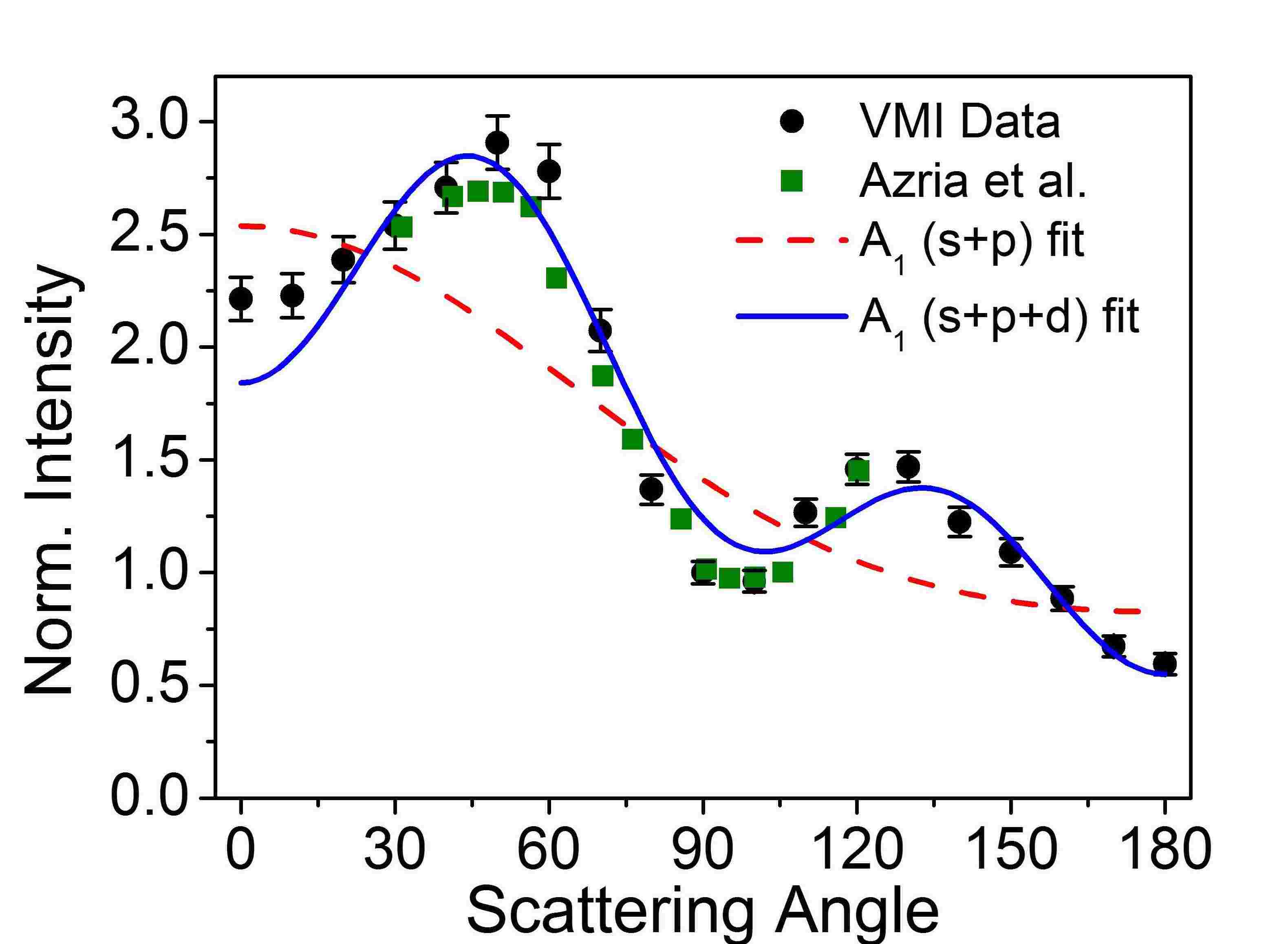}}
\caption{(a) Kinetic energy distribution of \ce{H-} ions across the third resonance. (b) Kinetic energy distribution of \ce{H-} ions at 7.5 eV fitted with Gaussian functions representing the spread of \ce{H-} kinetic energy due to vibrational and rotational excitation of SH fragment. The fit shows population of vibrational states upto $\nu$=4.  (c) Angular distribution of \ce{H-} ions at 7.5 eV compared with the data of Azria et al. \cite{c4azria} and fit with \ce{A1} symmetry functions. Fit with $s$, $p$ and $d$ waves matches the data very well.}
\label{fig4.7}
\end{figure}

The third resonance at 7.5 eV produces \ce{H-} ions dominantly as seen in ion yield curve in Figure \ref{fig4.1}. The maximum kinetic energy of \ce{H-} ions observed at this resonance is about 4 eV as seen in Figure \ref{fig4.7}(a). This indicates the dissociation channel to be \ce{H-} + SH (X $^{2}\Pi$) whose thermodynamic threshold is 3.15 eV. The maximum kinetic energy for this channel would be 4.3 eV which is close to our observed value of 4 eV. The kinetic energy distribution measured by Azria et al. \cite{c4azria} at 7.5 eV in $45^{\circ}$ degree direction showed SH fragments to be in $\nu$=0 state only. The kinetic energy distribution obtained in the present measurement given in Figure \ref{fig4.7}(a) is obtained by integrating over the entire $2\pi$ scattering range. Retrieving the SH vibrational state population by fitting Gaussian functions to the \ce{H-} kinetic energy distribution for electron energy of 7.5 eV revealed SH states upto $\nu$=4 being populated (see Figure \ref{fig4.7}(b)). The relative intensities of the vibrational levels are found to be in the ratio 1:0.31:0.22:0.17:0.07.

\begin{figure}[!htbp]
\centering
\subfloat[]{\includegraphics[width=0.33\columnwidth]{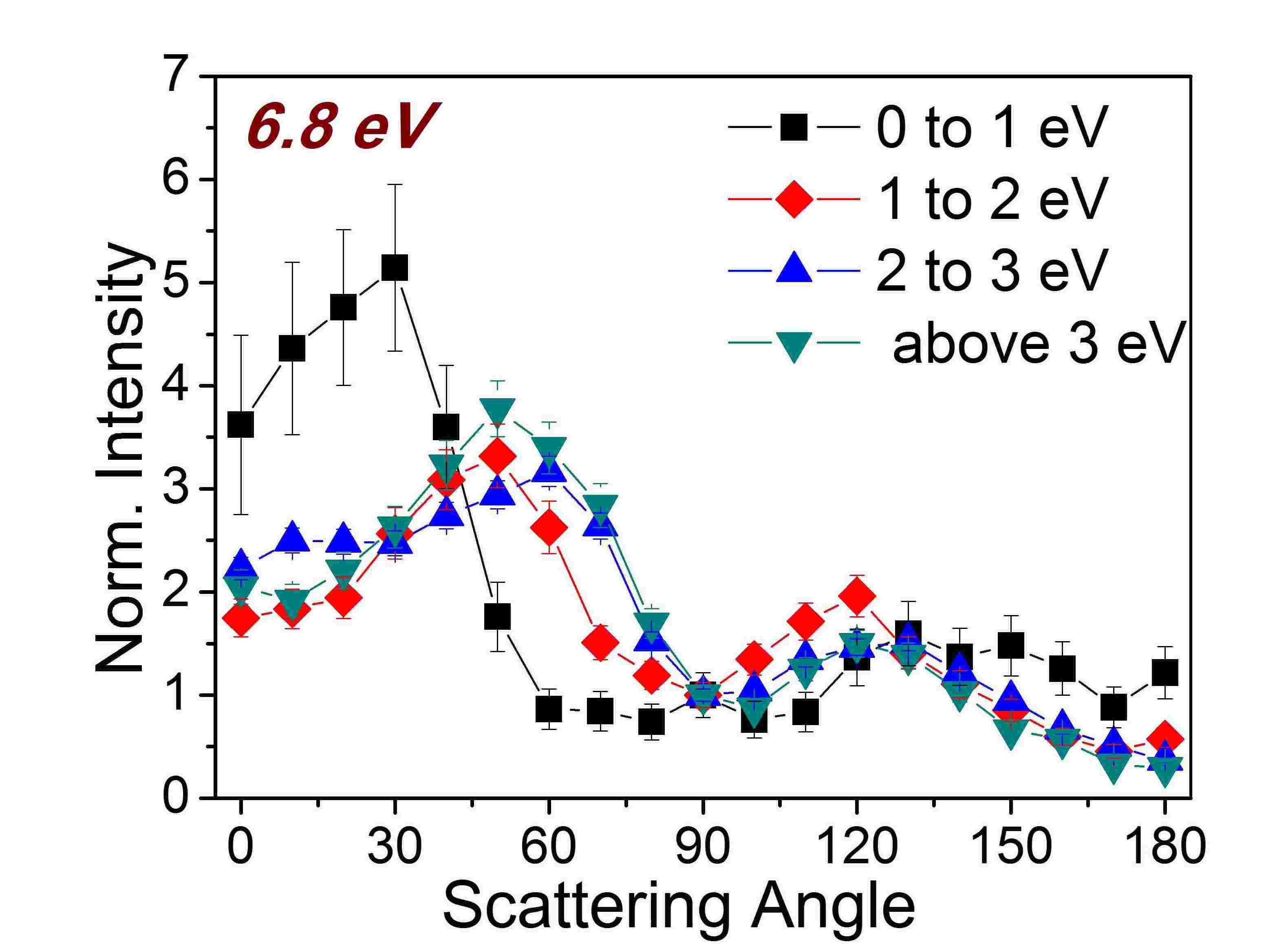}}
\subfloat[]{\includegraphics[width=0.33\columnwidth]{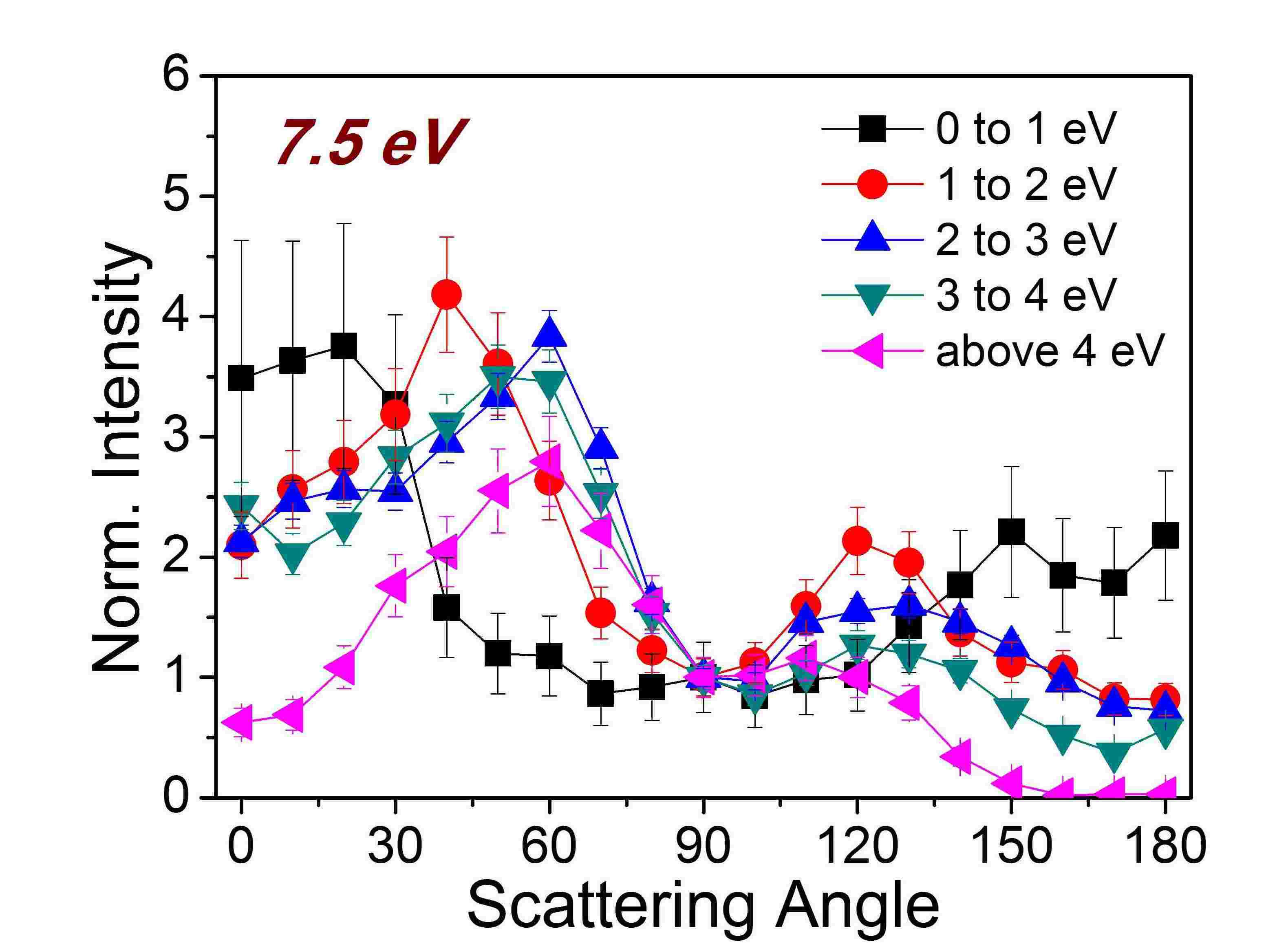}}
\subfloat[]{\includegraphics[width=0.33\columnwidth]{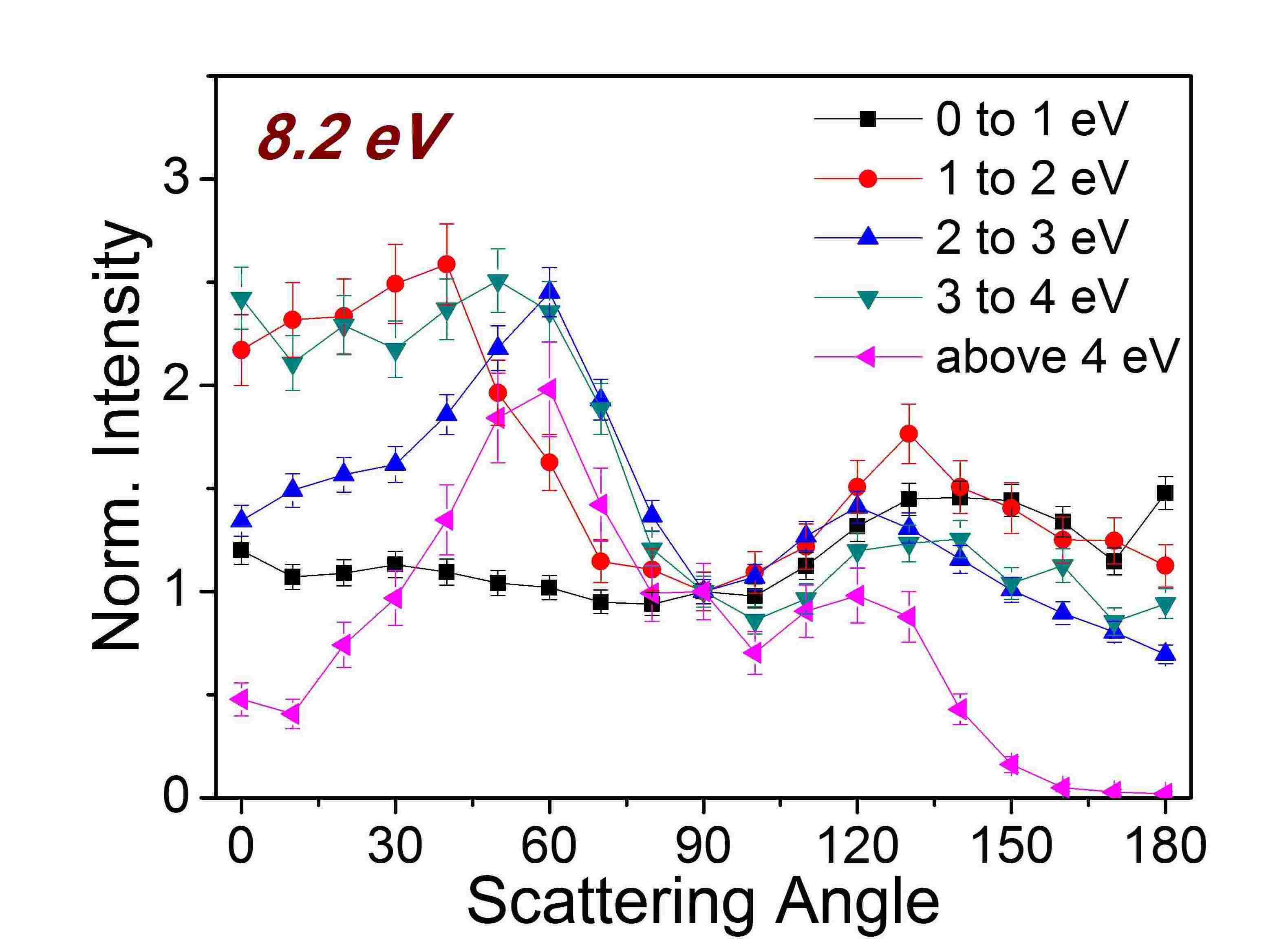}}
\caption{Angular distribution plots of \ce{H-} ions at 6.8 eV, 7.5 eV and 8.2 eV as a function of kinetic energy.}
\label{fig4.8}
\end{figure}

The angular distribution plot for \ce{H-} ions of all kinetic energies is given in Figure \ref{fig4.7}(c) along with the data of Azria et al \cite{c4azria}. Both the measurements show perfect agreement. The \ce{H2S^{-*}} anion symmetry state is \ce{A1} as seen by the fits in Figure \ref{fig4.7}(c) using \ce{A1} symmetry functions involving $s+p$ and $s+p+d$ partial waves. The relative amplitudes of the partial waves in the fits are found to be 1:0.85 with $\delta$=1.19 (for $s+p$) and 1:0.3:1.5 with $\delta_{1}$ (phase difference between $s$ and $p$ wave) and $\delta_{2}$ (phase difference between $p$ and $d$ wave) close to zero (for $s+p+d$). Further, we also looked for the variation of the angular distribution as a function of the kinetic energy (or internal excitation of SH fragment.) The angular plots as a function of KE are plotted in Figure \ref{fig4.8} for the incident electron energies 6.8 eV, 7.5 eV and 8.2 eV respectively. We see that the angular distributions show similar curves with peaks at $45^{\circ}$ and $135^{\circ}$ over the entire kinetic energy range. Unlike the angular distribution of \ce{H-} ions from the \ce{^{2}A1} resonance at 8.5 eV in water, where the angular distribution varies as a function of the vibrational state of OH, there is no such behaviour in \ce{H2S}. Thus, axial recoil approximation is found to hold good in the case of \ce{H2S} consistent with \ce{A1} symmetry in the entire fragmentation process at this resonance. That is, the molecular ion does not undergo any structural change like bending mode vibration during the dissociation process. Whereas in \ce{H2O}, the bending mode vibration of the molecule makes it linear leading to variation in the angular distribution of \ce{H-} ions with kinetic energy. Thus, we see intra-molecular vibrational redistribution (IVR) of excess energy into the bending mode oscillations to be dominant prior to the dissociation of the water anion whereas in \ce{H2S} there appear to be very little IVR.

\subsubsection{\ce{S-} ions}
Figures \ref{fig4.3} (e) and (f) show the velocity image of \ce{S-}/\ce{SH-} as a very small blob without any anisotropic structure and is attributed to the three body break up process \ce{S-} + H + H (threshold: 5.4 eV). The maximum KE estimated is about 0.12 eV. Energetically, \ce{SH-} produced via the \ce{SH-}($^{1}\Sigma^{+}$) + H (threshold: 1.6 eV) is also possible. At 7.5 eV, the excess energy available for this channel is 5.9 eV (more than the three body breakup threshold). Assuming that this excess energy is distributed as kinetic energy amongst the \ce{SH-} and H fragments, the maximum KE of \ce{SH-} would be about 0.18 eV. However, it is unlikely to be \ce{SH-} as there is sufficient energy available for three body break up and high resolution measurements by Abouaf and Billy \cite{c4abouaf} show only \ce{S-} at higher electron energies. No \ce{SH-} ion signal was seen in their measurements at electron energies beyond 4 eV. Also, the size of the velocity image is not commensurate with maximum KE expected from the  \ce{SH-}($^{1}\Sigma^{+}$) channel. This is similar to the case in \ce{H2O}, where the three body \ce{O-} + H + H (threshold 8.04 eV) channel is seen at the 8.5 eV resonance (\ce{^{2}A1} state) but no \ce{OH-} is seen. Thus, the puzzle of absence of \ce{OH-}($^{1}\Sigma$), even though favoured by an \ce{A1} resonance state seems to extend to \ce{H2S} as well where the \ce{SH-}($^{1}\Sigma$) appears to be absent at the \ce{^{2}A1} resonance about 7.5 eV. 

\subsection{Resonance process peaking at 9.6 eV}

There has been no report so far on the resonance at around 10 eV, though its existence has been seen in total ion yield measurements. Photo-electron/photo-absorption studies show the lowest excited states of \ce{H2S+} as of \ce{^{2}B1}, \ce{^{2}A1} and \ce{^{2}B2} symmetry in ascending order \cite{c4li,c4ibuki}. The first two are supposedly the grandparent states of the anion resonances at 5.2 and 7.4 eV respectively. Hence, the resonance at 9.6 eV may be understood as two Rydberg electrons attached to the \ce{^{2}B2} state of \ce{H2S+}.

\subsubsection{\ce{H-} ions} 

\begin{figure}
\centering
\subfloat[]{\includegraphics[width=0.4\columnwidth]{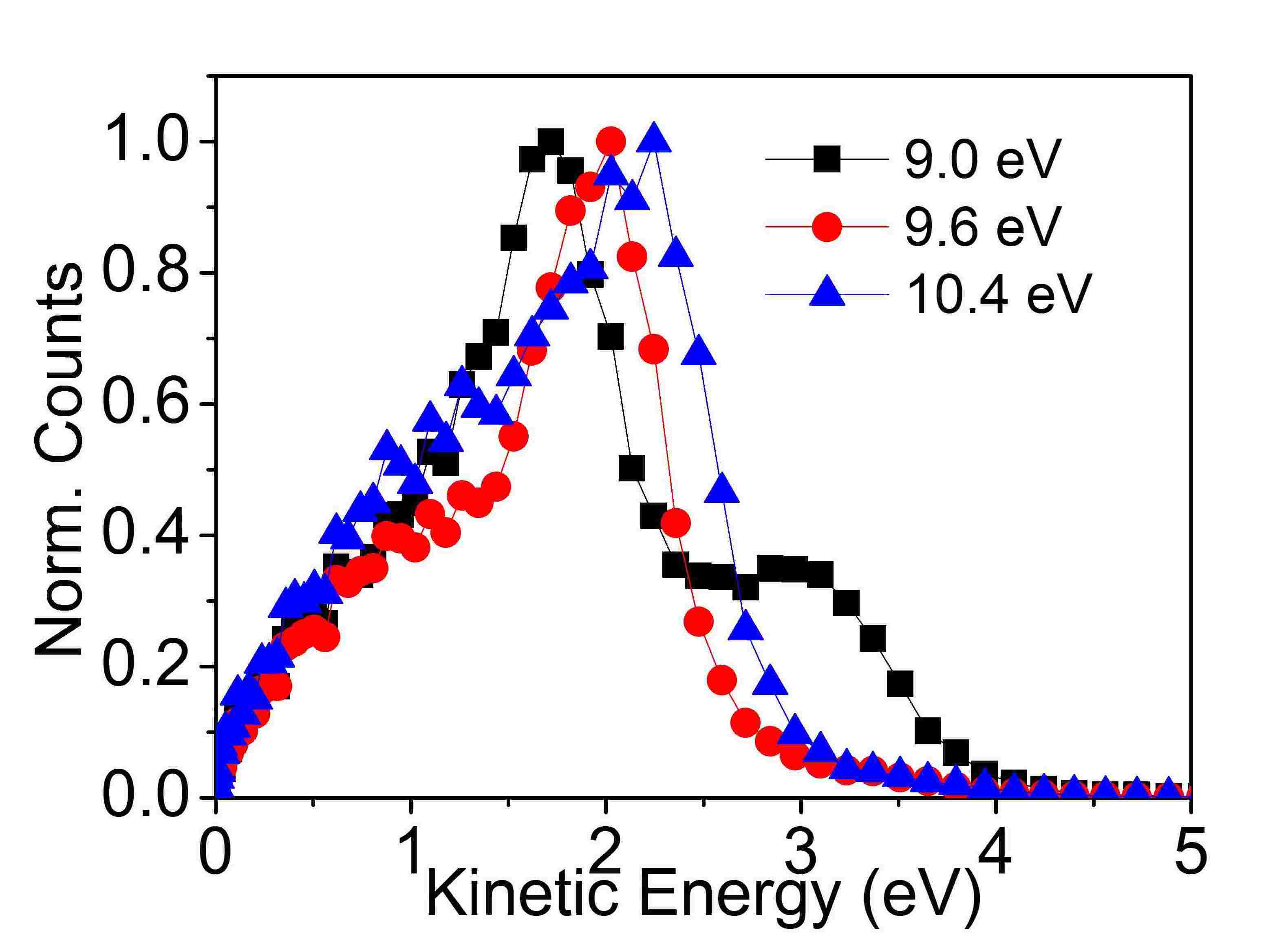}}
\subfloat[]{\includegraphics[width=0.4\columnwidth]{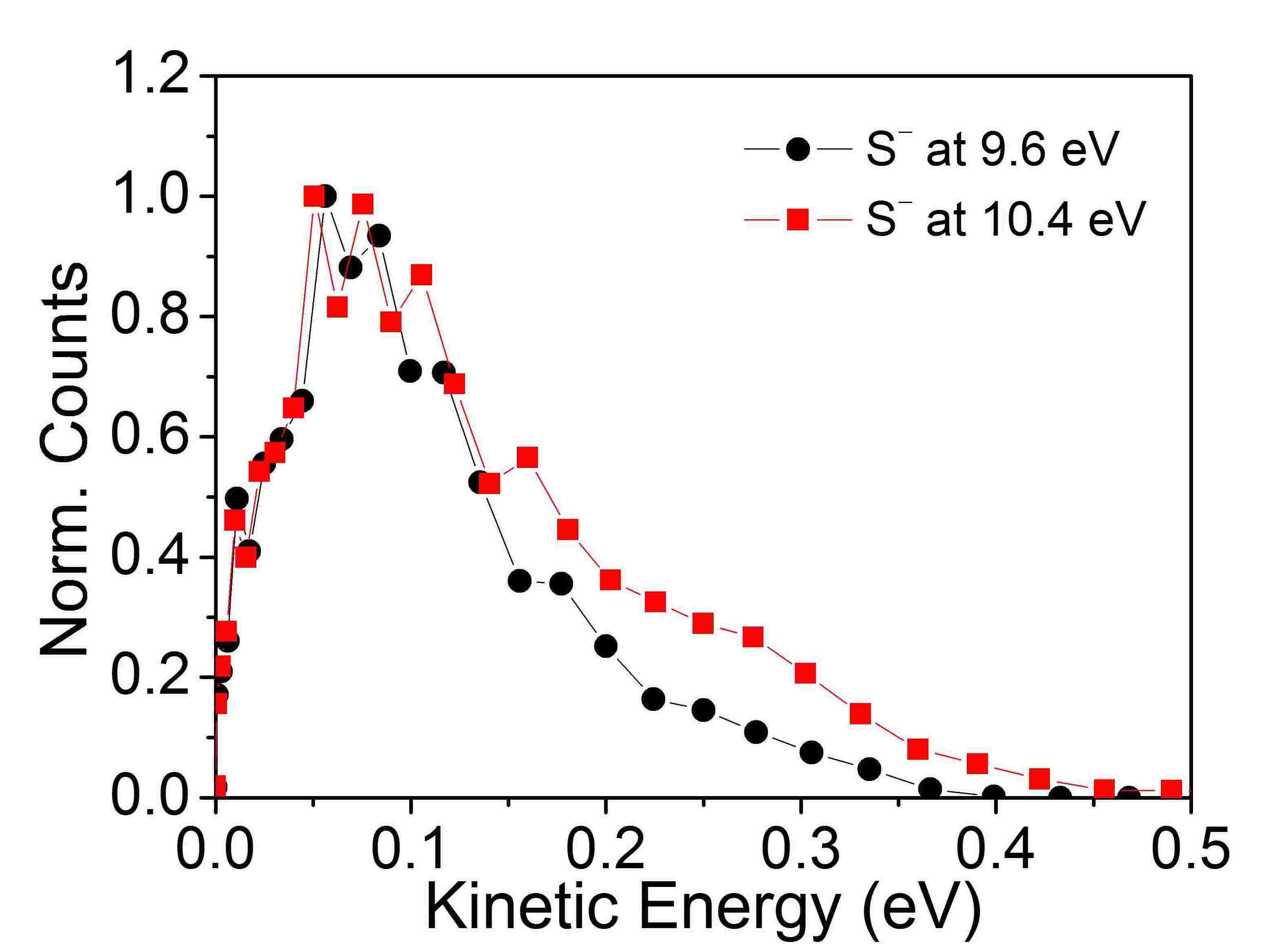}}
\caption{(a) Kinetic energy distribution of \ce{H-} ions at 9.0 eV, 9.6 eV and 10.4 eV (b) Kinetic energy distribution of \ce{S-} ions at 9.6 eV and 10.4 eV}
\label{fig4.9}
\end{figure}

A comparison of the resonances in \ce{H2O} and \ce{H2S} indicate that the DEA process in \ce{H2S} centered at 9.6 eV is similar to the 11.8 eV resonance in \ce{H2O}. The \ce{H2S} at this resonance produces \ce{H-} ions scattered in forward and backward angles in three groups (inner, middle and outer) as shown in the velocity map images (see Figure \ref{fig4.2}-(g),(h),(i)). The kinetic energy distribution of \ce{H-} ions for electron energies 9.0 eV, 9.6 eV and 10.4 eV plotted in Figure \ref{fig4.9}(a). The plot for 9.0 eV shows clearly three structures - first between 0 and 1.5 eV , second between 1.5 eV and 3 eV and the third one beyond 3 eV. While the outer structure beyond 3 eV is seen clearly at 9 eV, it becomes weaker as the electron energy increases. i.e. at 9.6 eV and 10.4 eV. The structure in the kinetic energy spectrum beyond 3 eV is attributed to the \ce{H-} + SH ($X ^{2}\Pi$) channel with threshold energy of 3.15 eV. At incident electron energy of 9.6 eV, the maximum kinetic energy of \ce{H-} is estimated to be about 6.5 eV. However, we do not observe ions with energy more than 4 eV indicating that SH is being formed in very high vibrational states, close to its dissociation limit. The structure between 1.5 and 3 eV with a peak at 2 eV points to \ce{H- + SH^{*} ($A ^{2}\Sigma$)} dissociation channel where the neutral SH fragment is in the first electronic excited state. The peak in the kinetic energy spectrum shows that this is the dominant fragmentation channel of the resonance. This may appear a trifle surprising since the threshold for this channel at 6.9 eV is higher than the threshold (6.74 eV) for three-body fragmentation channel. However, we note that the dissociation limit for \ce{SH^{*}} (A $^{2}\Sigma$) is \ce{S}(\ce{^{1}D}) + H, which is 1.15 eV above S(\ce{^{3}P}) + H \cite{c4nist,c4hirst}. Using the threshold energy (6.9 eV) and the dissociation limit into \ce{H-} + H + S(\ce{^{1}D}) (7.89 eV), the kinetic energy of \ce{H-} would be in the range of 1.7 to 2.62 eV taking incident electron energy to be 9.6 eV. This appears to be in reasonable agreement with our observation. The structure between 0 to 1.5 eV indicates the three body breakup channel \ce{H-} + H + S. This process has threshold energy of 6.74 eV. For the instantaneous symmetric three body breakup of the resonant state, the kinetic energies of \ce{H-} and \ce{S-} ions as a function of half the bond angle H-S-H ($\theta$) are given by 

\begin{eqnarray}
E_{H^{-}} = \frac{8 E_{o}}{16 + \cos^{2}\theta} \\
E_{S^{-}} = \frac{E_{o}\cos^{2}\theta}{16 + \cos^{2}\theta}
\end{eqnarray}

where $E_{o}$ is the total kinetic energy release. At the electron energy of 9.6 eV, $E_{o}$ is 2.86 eV for the \ce{H-} channel and 4.2 eV for the \ce{S-} channel.  Under axial recoil approximation, i.e., for the \ce{H2S} equilibrium bond angle of $95^{\circ}$, \ce{H-} will have a kinetic energy equal to 1.39 eV. If we assume that the molecule undergoes bending motion during the fragmentation, assuming symmetric fragmentation of the two S-H bonds, the minimum and maximum kinetic energies are found to be 1.34 eV ($0^{\circ}$) and 1.43 eV ($180^{\circ}$) respectively. However, this does not explain the kinetic energy distribution seen below 1.3 eV and extending down to thermal energies. There are two possible ways this may happen. S atom has electronic excited states, \ce{^{1}D} and \ce{^{1}S} which are 1.15 eV and 2.75 eV above the ground state \cite{c4nist, c4hirst}. If the S atom is formed in either of these states, the excess kinetic energy would be correspondingly reduced to give the \ce{H-} kinetic energy from an instantaneous three body break as 0.83 eV and 0.05 eV respectively. The second possibility is for the fragmentation to occur in a sequential process through an intermediate \ce{SH-} state. In this case the H atom could take away most of the excess kinetic energy. The subsequent fragmentation of \ce{SH-} could produce \ce{H-} with low kinetic energies below 1.3 eV. We have observed a similar behaviour of the kinetic energy spectrum of \ce{H-} in the 12 eV resonance in water. There we could clearly identify the low kinetic energy release as due to a sequential fragmentation through an \ce{OH-} state and rule out the production of excited O atoms. Here in the case of \ce{H2S}, we are unable to distinguish the two possibilities. 

\subsubsection{$S^{-}$ ions}
The $S^{-}$ velocity images at 9.6 eV and 10.4 eV - Figures \ref{fig4.3}(g) and (h) show a distinct forward-backward ring along with a small central blob with near zero energy. These appear to be superimposed on a fairly uniform intensity distribution. The kinetic energy distributions of \ce{S-} in Figure \ref{fig4.9}(b) peak below 0.1 eV with tails extending beyond 0.3 eV. Using equation 4.2 above, we can see that the kinetic energy distribution expected for \ce{S-} via the \ce{S-} + H + H channel (threshold: 5.4 eV) at 9.6 eV and 10.4 eV electron energy range from 0 to 0.25 eV and 0 to 0.29 eV respectively assuming an instantaneous three body breakup process. The tails we see extending beyond these energy ranges are from the imaging process. The forward-backward rings of $S^{-}$ result from a dissociation process where the bond angle of \ce{H2S} decreases from $95^{\circ}$ to close to $0^{\circ}$ upon electron attachment and consequently, the \ce{S-} is ejected out either in forward or backward angles with respect to the electron beam. The central blob in the velocity image with close to zero energy may correspond to the formation of \ce{S-} when the bond angle increases to $180^{\circ}$ or through the sequential fragmentation process \ce{H2S^{-*} -> H + SH^{-*} -> H + H + S-}  where it is left with near zero energy.

\subsubsection{Angular Distributions}
\begin{figure}[!htbp]
\centering
\subfloat[]{\includegraphics[width=0.4\columnwidth]{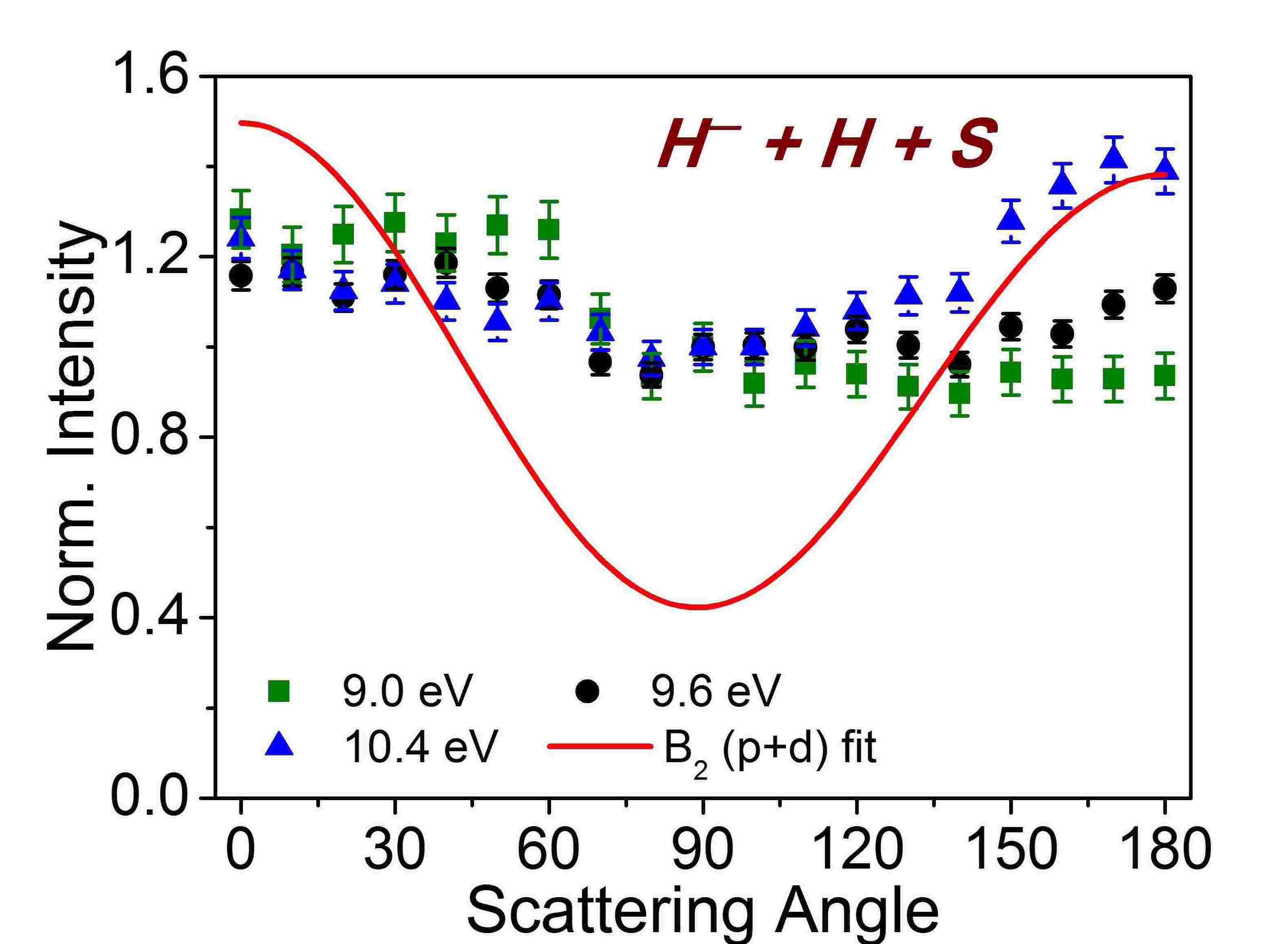}}
\subfloat[]{\includegraphics[width=0.4\columnwidth]{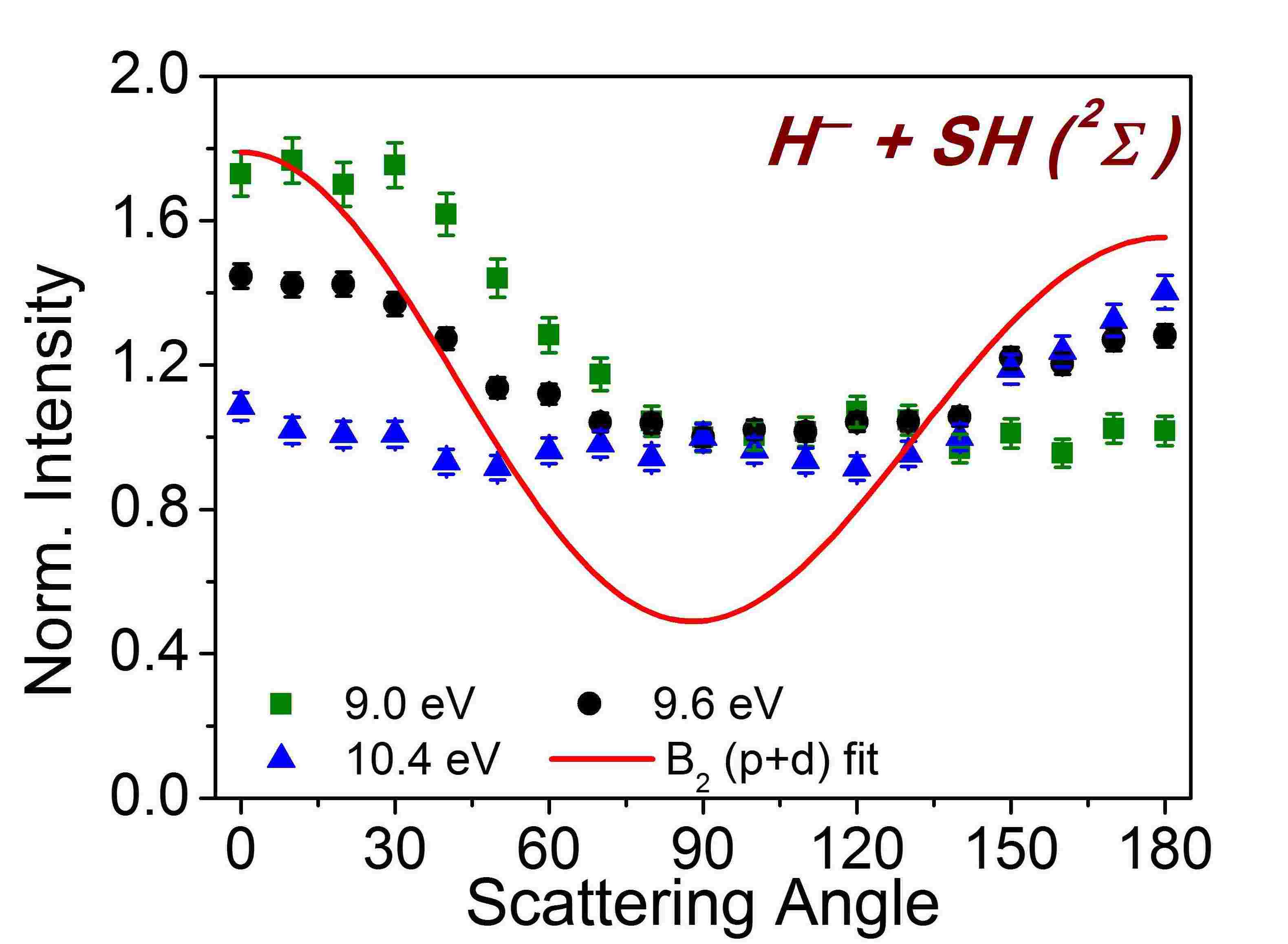}}\\
\subfloat[]{\includegraphics[width=0.4\columnwidth]{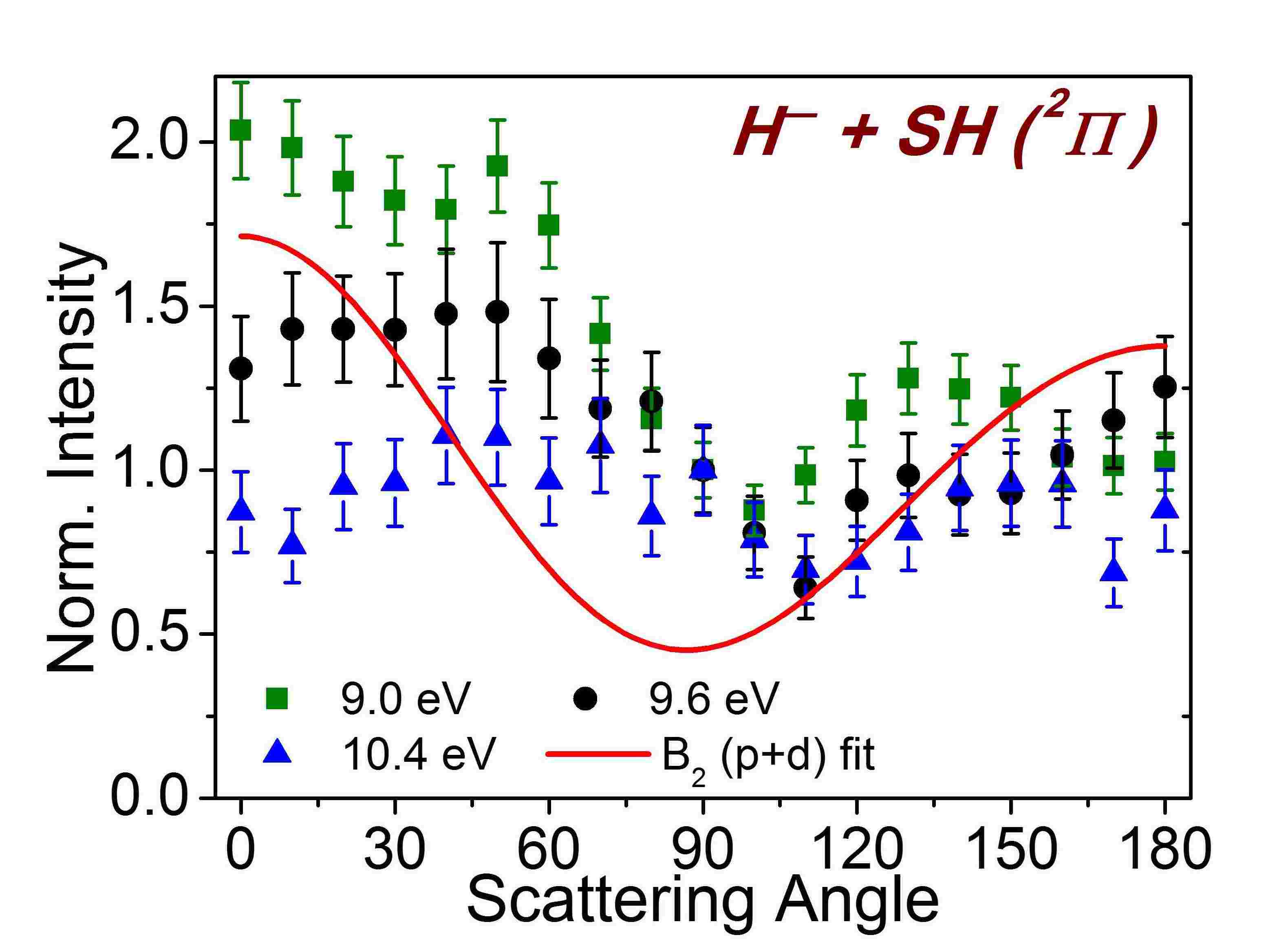}}
\subfloat[]{\includegraphics[width=0.4\columnwidth]{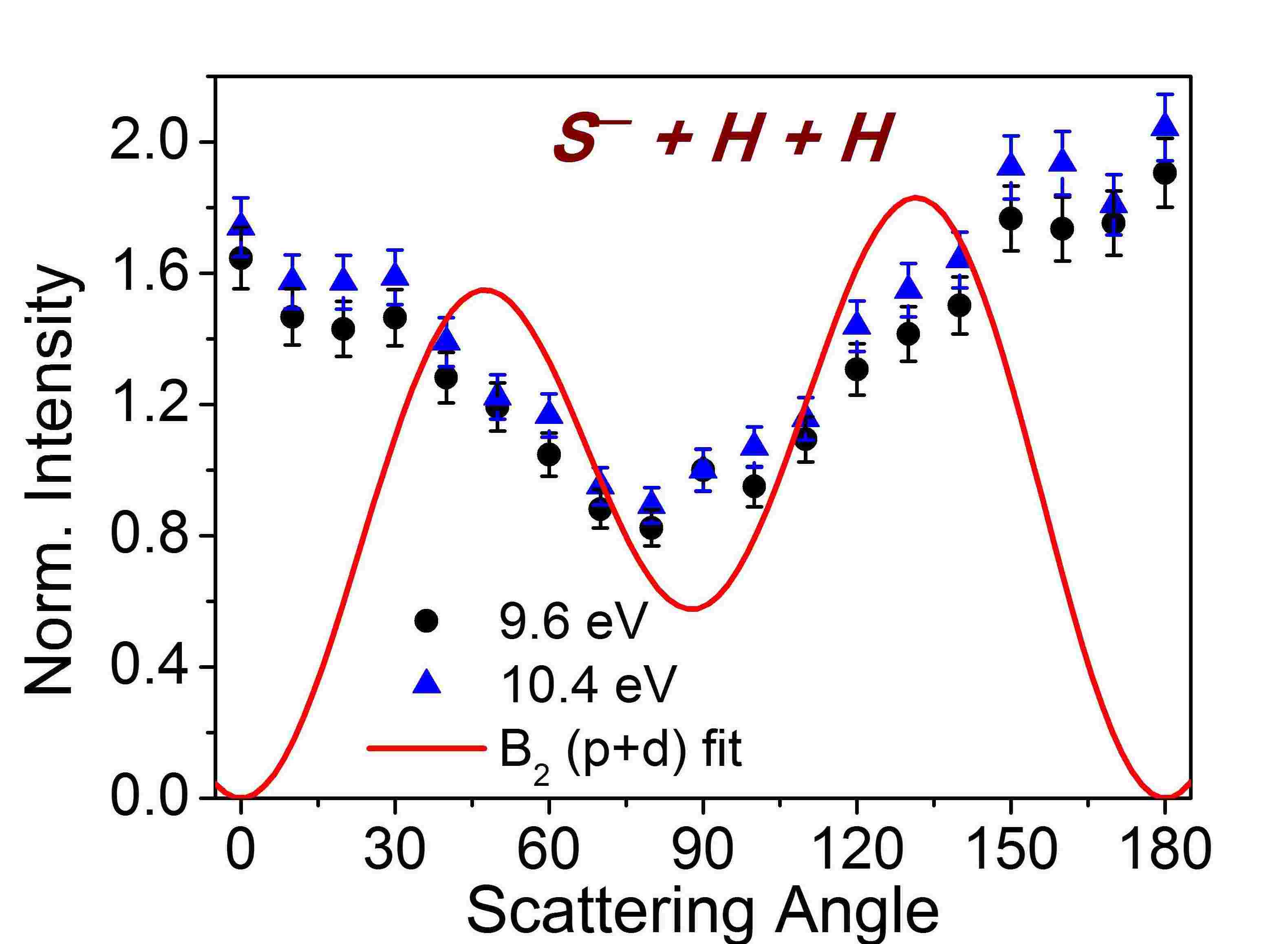}}
\caption{Angular distributions of \ce{H-} ions produced via (a) \ce{H-} + H + S (b) \ce{H-} + SH($^{2}\Sigma$) (c)\ce{H-} + SH ($^{2}\Pi$) at 9.0, 9.6 and 10.4 eV electron energies. (d) Angular distribution of \ce{S-} ions produced via the \ce{S-} + H + H channel at 9.6 and 10.4 eV. The red curve in each of these plots is the fit obtained using \ce{B2} symmetry functions ($p$ and $d$ partial waves) under axial recoil approximation. The fits qualitatively reproduce the features in the angular distribution data but not exactly. This suggests change in the molecular geometry upon electron attachment leading to angular distribution not described by the axial recoil based fits.}
\label{fig4.10}
\end{figure}

The angular distribution of \ce{H-} and \ce{S-} ions produced via various dissociation channels across the resonance at 9.0 eV, 9.6 eV and 10.4 eV are plotted in Figure \ref{fig4.10}. The red curves in each of the plots are fits obtained using the \ce{B2} symmetry function taking $p$ and $d$ partial waves and fit to the 9.6 eV data. The fits in Figure \ref{fig4.10}(a), (b) and (c) are for \ce{H-} ions and obtained by assuming the dissociation of one of the SH bonds (oriented at $47.5^{\circ}$ with respect to the molecular symmetry axis) in ground state equilibrium geometry of the neutral molecule. Whereas the fit in Figure \ref{fig4.10}(d) is for \ce{S-} ions obtained assuming that the ion is ejected along \ce{C2} axis. Hence, the dissociation axis is same as the molecular symmetry axis (i.e. $0^{\circ}$) and this is taken into account while fitting the data with the \ce{B2} symmetry curves. In case of \ce{H-} ions from the three channels, the fit is predominantly due to a $p$-wave component with very little $d$-wave contribution ($p:d$=1:0.05 and  $\delta$ $\sim$ 0). The fit qualitatively reproduces the forward-backward distribution with a dip in between but is far from being a good fit and thus, suggests that the dissociation is not describable by the axial recoil based fits. Similar is the case for \ce{S-}, where the only resemblance between the data and the fit is the dip about the $90^{\circ}$, but seriously in disagreement at forward-backward angles. The ratio of $p$ to $d$ partial waves is 1:1.8 with a phase difference of 1.7 radians. We believe that the major contribution to the deviation from axial recoil approximation is the bending mode vibrations in the molecule. Such a process has been seen at the 8.5 eV and 12 eV resonances in water.

\subsection{Comparison with \ce{B2} resonance in water}

\begin{figure}[!h]
\centering
\subfloat[\ce{H- + H + S} (blue) and \ce{H- + H + O} (red)]{\includegraphics[width=0.4\columnwidth]{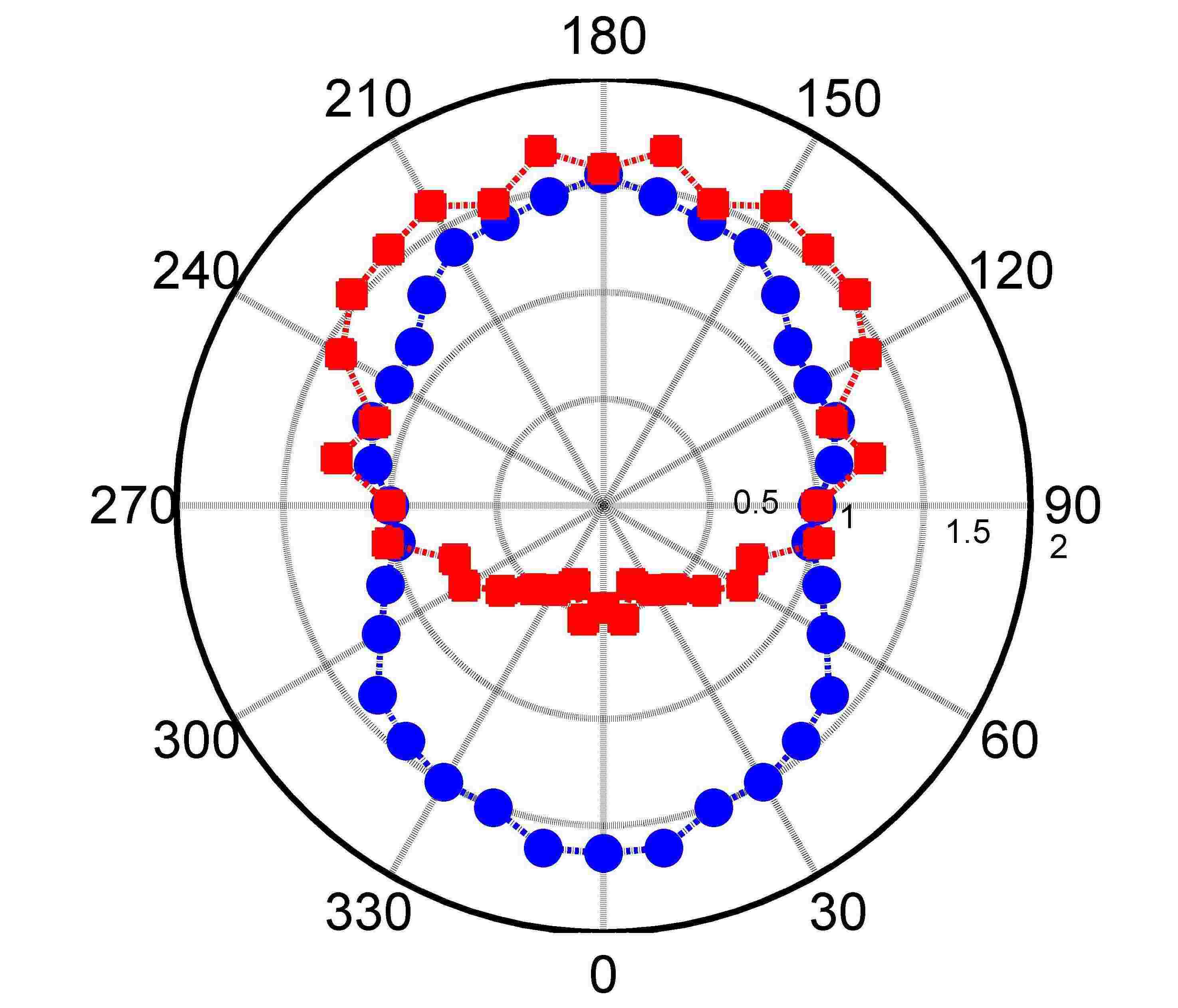}} \hspace{1cm}
\subfloat[\ce{H-} + SH ($^{2}\Sigma$)(blue) and \ce{H-} + OH ($^{2}\Sigma$)(red)]{\includegraphics[width=0.4\columnwidth]{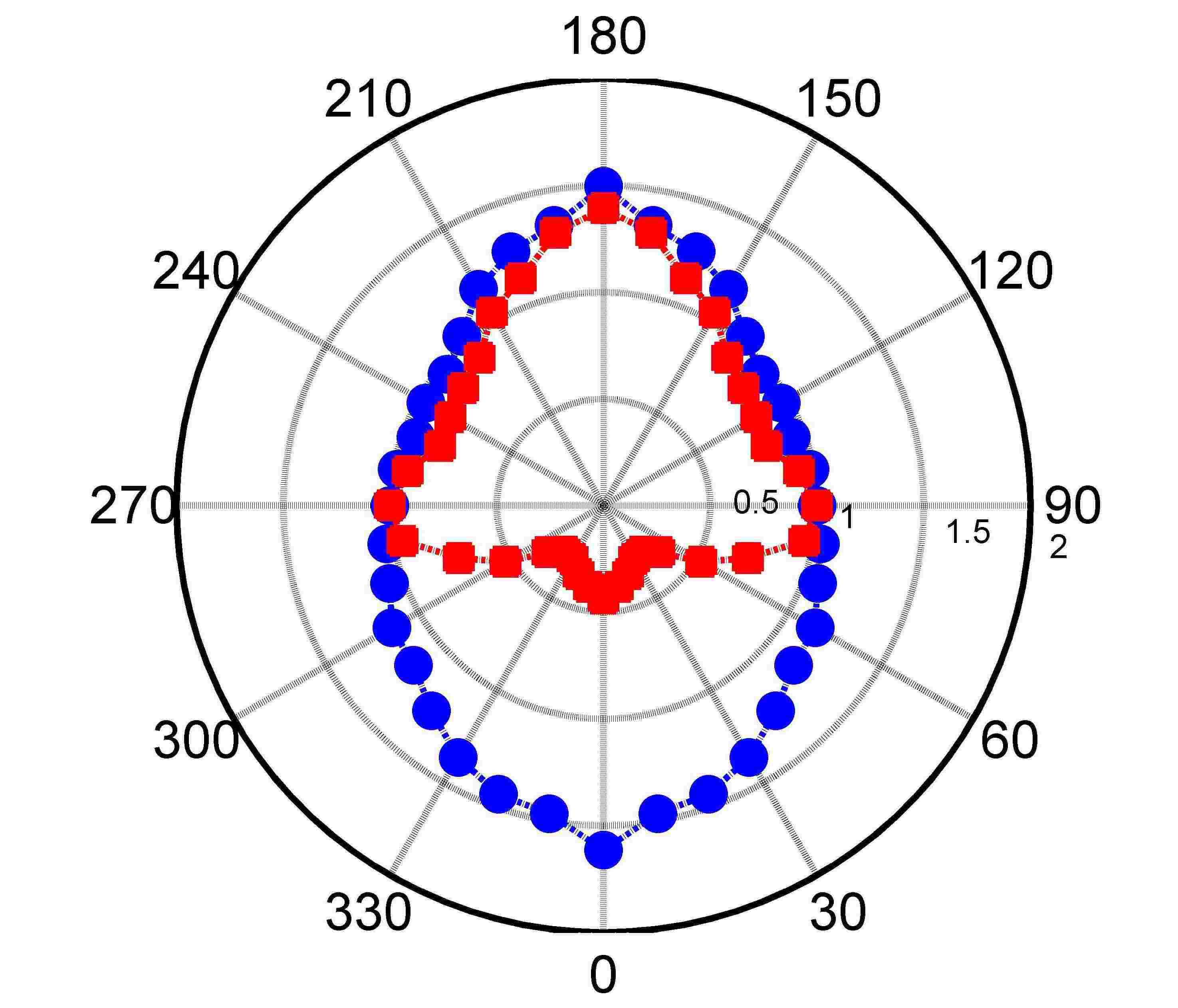}}\\
\subfloat[\ce{H-} + SH ($^{2}\Pi$)(blue) and \ce{H-} + OH ($^{2}\Pi$)(red)]{\includegraphics[width=0.4\columnwidth]{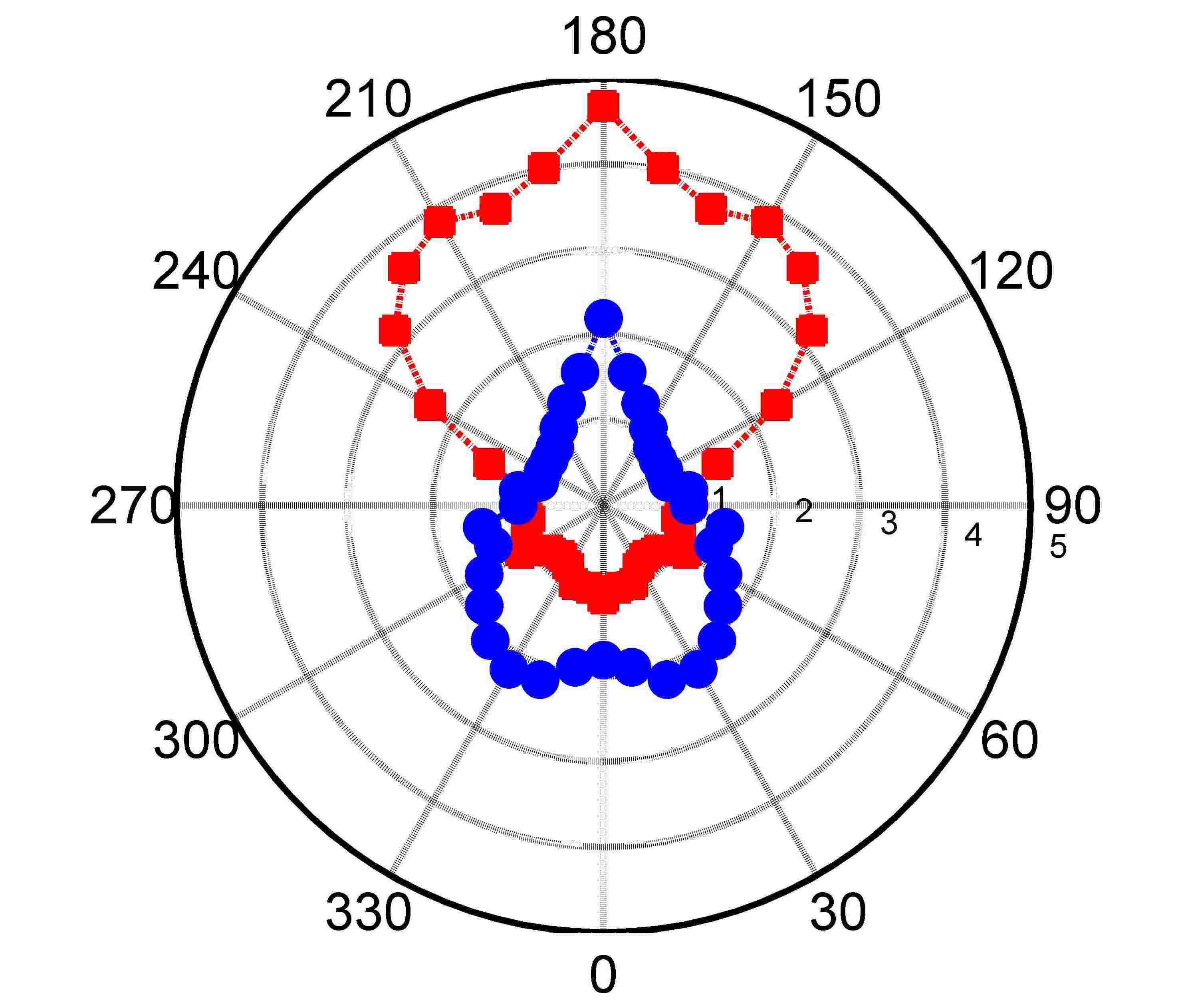}} \hspace{1cm}
\subfloat[\ce{S- + H + H} (blue) and \ce{O- + H + H} (red)]{\includegraphics[width=0.4\columnwidth]{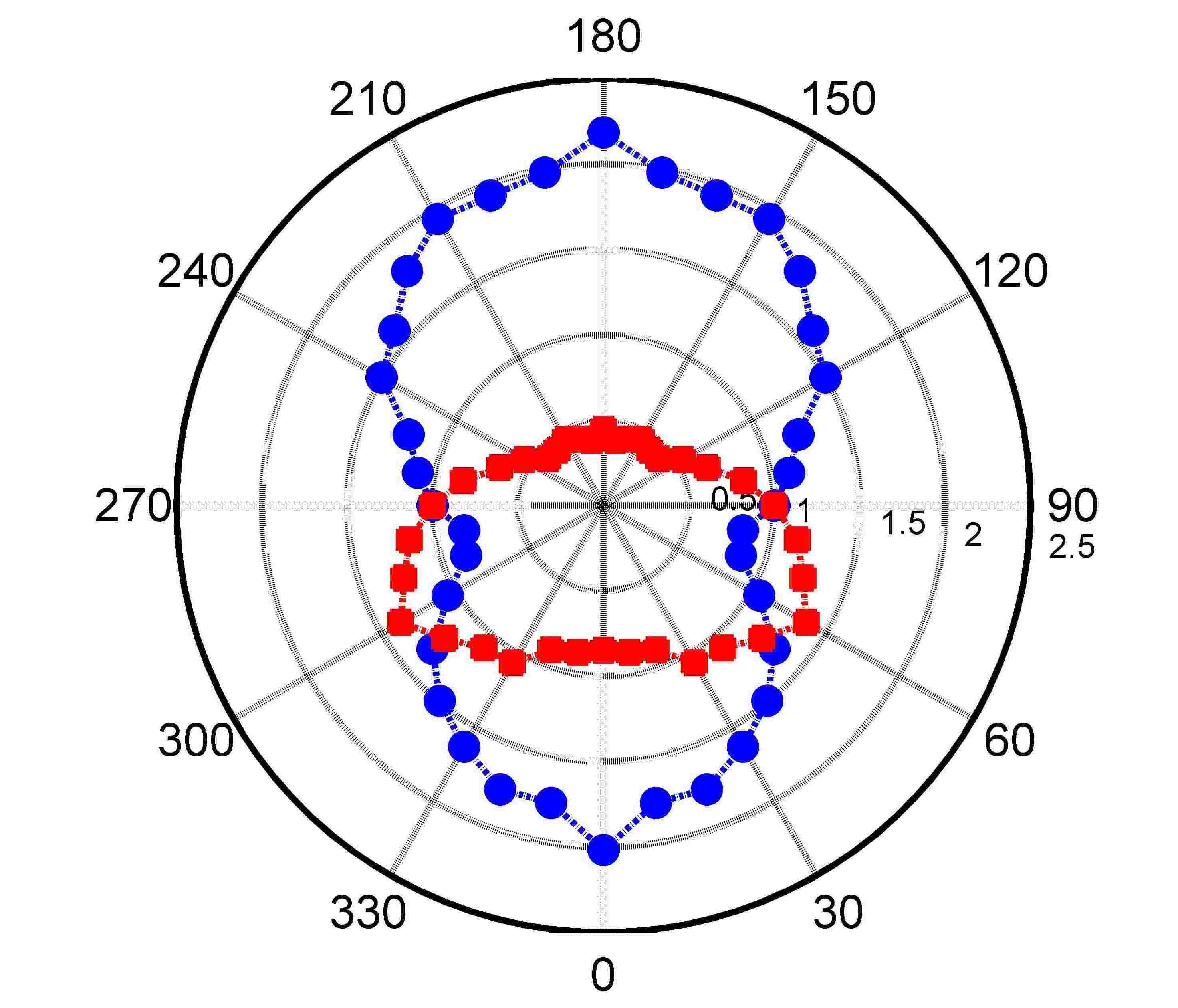}}
\caption{Comparison of the angular distributions of \ce{H-}/\ce{H2O} with \ce{H-}/\ce{H2S} and \ce{O-}/\ce{H2O} with \ce{S-}/\ce{H2S} produced in similar dissociation channels from the \ce{B2} resonance in these molecules. Blue circles are used for \ce{H2S} and red squares are for \ce{H2O}.}
\label{fig4.11}
\end{figure}

As mentioned earlier, the overall kinematics of the 9.6 eV resonance in \ce{H2S} is very similar to that of the 11.8 eV resonance in \ce{H2O}. This includes the three possible dissociation channels as well as the sequential process in the three-body break up channel with \ce{SH-} and \ce{OH-} as the respective intermediate species. A comparison of the angular distributions of fragment ions in \ce{H2S} and \ce{H2O} shows two distinct features (see Figure \ref{fig4.11}). In the case of water the angular distributions of \ce{H-} and \ce{O-} showed distinct forward-backward anisotropy with \ce{H-} being ejected strongly in the backward direction while \ce{O-} being ejected predominantly in the forward direction. These angular distributions are found to be independent of electron energy across the resonance. In contrast to the case of \ce{H2O}, the angular distribution of the \ce{H-} fragment ions from \ce{H2S} at this resonance appear to be in better conformity with what is expected of a \ce{B2} resonance as seen in Figure \ref{fig4.10}. We also note that unlike the case of \ce{H2O}, where the \ce{H-} from the three dissociation channels appear to have fairly different angular distributions, that from \ce{H2S} seems to have relatively similar angular distributions in all the three channels. Also, the angular distribution of \ce{H-} from \ce{H2S} shows strong energy dependence across the resonance as seen clearly in the \ce{H-} + SH ($^{2}\Sigma$) channel angular distribution plots in Figure \ref{fig4.10}(b). At 9 eV the forward angles have more intensity while at 10.4 eV the backward angles have more intensity. At 9.6 eV, the intensity distribution appears to be more or less symmetric about $90^{\circ}$. 

In conclusion, the 9.6 eV resonance in \ce{H2S} and the 11.8 eV resonance in \ce{H2O} are found to be very similar in the dissociation kinematics. Based on the kinetic energy distribution, we find that the formation of \ce{H-} from these resonances occurs through three different channels leading to the formation of SH (OH) in the ground state ($^{2}\Pi$), SH (OH) in the first electronically excited state ($^{2}\Sigma$) and the three-body fragmentation. It is also seen that the three-body fragmentation has a contribution arising from a sequential process through an intermediate \ce{SH^{-*}} (\ce{OH^{-*}}) state. While the dissociation pathways seem very similar, the angular distributions of the fragments show contrasting behaviour. In the case of \ce{H2O}, the \ce{H-} and \ce{O-} are preferentially emitted in opposite hemispheres, while in the case of \ce{H2S} we do not observe such strong asymmetry.  The fits for the angular distribution data using the axial recoil approximation agrees reasonably well with a \ce{B2} symmetry in the case of \ce{H2S}, while the unusual angular distribution in \ce{H2O} show considerable deviation. We also note that the angular distributions from \ce{H2O} in all the channels are relatively independent of electron energy, while in \ce{H2S} it appears to be quite sensitive to the electron energy across the resonance. Though we are unable to explain the observed differences, it appears that the answers have to be sought in the comparatively large size of the S atom and the presence of the d-electrons. 

\section{Summary}

\begin{enumerate}

\item	Kinetic energy and angular distribution of \ce{H-} and \ce{S-}/\ce{SH-}  ions from DEA to \ce{H2S} obtained in the 1-10 eV electron energy range using VMI technique and comparison with results on \ce{H2O}.

\item	First resonance centered at 2.4 eV 
\begin{enumerate}
\item	Shape resonance producing \ce{S-} and \ce{SH-} ions mostly. 
\end{enumerate}

\item	Second resonance centered at 5.2 eV 
\begin{enumerate}
\item Angular distribution measurements on \ce{H-} in agreement with Azria et al. \cite{c4azria} 
\item Confirms the dynamics due to \ce{(2b1)^{-1} (6a1)^{2} -> ^{2}B1} resonance.
\item	\ce{H-} + SH ($^{2}\Pi$) and \ce{S- + H2} channels present similar to the first resonance process in water at 6.5 eV.
\item	\ce{SH ($^{2}\Pi$)} internal excitation less intense as compared to \ce{OH (^{2}$\Pi$)} in water at 6.5 eV.
\end{enumerate}

\item	Third resonance centered at 7.5 eV 
\begin{enumerate}
\item Angular measurements on \ce{H-} in agreement with Azria et al \cite{c4azria} confirming \ce{(5a1)^{-1} (6a1)^{2} -> ^{2}A1} resonance. 
\item	Angular distribution independent of kinetic energy unlike the case of 8.5 eV resonance in water, showing very little intra-vibrational redistribution. 
\item	\ce{SH- ($^{1}\Sigma$)} fragment not seen at this resonance. Instead, the three body channel \ce{S- + H + H} present.
\end{enumerate}

\item Fourth resonance centered at 9.6 eV 
\begin{enumerate}
\item	First report of kinetic energy and angular distribution of \ce{H-} and \ce{S-} ions at this resonance.
\item	\ce{^{2}B2} resonance symmetry - similar to 11.8 eV resonance in water.
\item Four dissociation channels seen - \ce{H- + H + S}, \ce{H- + SH ($^{2}\Sigma$)}, \ce{H- + SH($^{2}\Pi$)} and \ce{S- + H + H}
\item	Dissociation channels and kinematics similar to water, but different angular distributions.
\end{enumerate}

\end{enumerate}

\end{document}